\documentclass[12pt]{article} 
\usepackage[a4paper,textwidth=16cm,textheight=23cm,footnotesep=\baselineskip]{geometry}
\usepackage[english]{babel}
\usepackage{amsmath}
\numberwithin{equation}{section}
\usepackage{amssymb} 
\usepackage{graphicx} 
\usepackage{float} 
\usepackage{hyperref}
\usepackage[all]{hypcap}
\usepackage{bm} 
\usepackage{slashed}
\usepackage[utf8]{inputenc}
\DeclareSymbolFont{matha}{OML}{txmi}{m}{it}
\DeclareMathSymbol{\varv}{\mathord}{matha}{118} 
\let\vec\mathbf 
\newcommand{\abs}[1]{\left| #1 \right|} 
\newcommand{\avg}[1]{\left< #1 \right>} 
\usepackage{cite} 
\usepackage{xcolor}
\usepackage{hyperref}
\usepackage[all]{hypcap}
\usepackage[T1]{fontenc} 
\hypersetup{
    colorlinks,
    linkcolor={red!50!black},
    citecolor={blue!50!black},
    urlcolor={blue!80!black}
			}

\newcommand{\spline}{spline}

\renewcommand{\varv}{v}
\begin{document}
\setlength{\baselineskip}{0.6cm}
\begin{centering}

  \textbf{\Large{ 
      Sterile neutrino dark matter:
	 \\[2.5mm]
      Impact of active-neutrino opacities
    }}

\vspace*{.6cm}

Dietrich B\"odeker
\footnote{bodeker@physik.uni-bielefeld.de} 
and Alexander Klaus
\footnote{aklaus@physik.uni-bielefeld.de}

\vspace*{.6cm} 

{\it Fakult\"at f\"ur Physik, Universit\"at Bielefeld, 33501 Bielefeld, Germany}

\vspace{8mm}

\end{centering}

\begin{abstract}
\noindent
The resonant production of keV sterile-neutrino dark matter mainly
takes place during the QCD epoch of the early universe.  
It has been argued that  it could be strongly 
affected by the opacities (or damping rates)
of active neutrinos, which receive non-perturbative QCD-contributions. 
We find that for lepton asymmetries 
$ n _ { L _ \alpha  } /s $ below $ 10 ^ { -6 } $
the opacities significantly affect the sterile-neutrino yield, but that
for larger asymmetries, which are necessary for producing a significant
fraction of the dark matter, the yield is  
insensitive to changes of the opacities.
Thus non-perturbative QCD
contributions to the opacities at temperatures around 160 MeV
will not affect this dark matter scenario. 
We obtain larger sterile-neutrino yields  than previous studies,
and thus 
weaker lower limits on the active-sterile mixing angle from Big Bang Nucleosynthesis.
\end{abstract}

\section{Introduction}	\label{s:intro}

Sterile neutrinos with mass in the keV range have long been discussed 
a dark matter candidate~\cite{Dodelson:1993je}.
Through their mixing with active neutrinos they can be produced
from the Standard Model (SM) plasma in the early  universe.
They are warm dark matter, giving rise to less structure on small scales
than cold dark matter, which may solve some problems of the 
standard $ \Lambda $CDM model. 
The detection  of small scale structure
in the Lyman-$ \alpha  $ forest 
results in lower bounds on the mass of the sterile neutrinos~\cite{Viel:2005qj}.
The sterile neutrinos can decay into an active neutrino and a photon,
giving rise to monochromatic X-rays. Non-observations of X-ray
lines imposes upper limits to the active-sterile mixing angle which 
depend on the sterile-neutrino mass.
The original production scenario proposed by Dodelson and Widrow 
\cite{Dodelson:1993je}, 
has already been ruled out as the sole source of dark matter production by
combining Lyman-$\alpha$ and X-ray constraints
\cite{Seljak:2006qw,Boyarsky:2009ix,Boyarsky:2008xj}.
One way to circumvent these constraints was suggested by Shi and Fuller 
\cite{Shi:1998km}:
They assume  lepton asymmetries much larger than
the observed baryon asymmetry. They lead to resonant  production, 
resulting generally in a non-thermal spectrum, 
which can be colder than in the Dodelson-Widrow scenario
(see, however,~\cite{Schneider:2016uqi}), and so
evade Lyman-$\alpha$ constraints.
Furthermore, resonant enhancement makes the production much  more efficient, 
requiring smaller mixing angles and thus escaping the upper limits from 
X-ray constraints
\cite{Abazajian:2001nj, Laine:2008pg,Essig:2013goa}. 

In order to calculate the final abundances, one has to track a set of 
coupled evolution equations 
for the sterile-neutrino phase space densities and the lepton asymmetries, 
the latter of which get depleted during the resonant conversion process.  
One difficulty in solving these equations comes from high lepton asymmetries 
resulting in sharp resonances, which necessitate high numerical precision. 
Furthermore, the production process typically starts at temperatures of a 
few GeV and ends at a few MeV prior to the onset of 
Big Bang Nucleosynthesis (BBN), 
introducing uncertainties from the QCD epoch 
$ T  \sim 160 $~MeV, when QCD interactions are strong
and thus non-perturbative.

The cosmological expansion depends on the pressure-energy relation
(equation of state) of the matter in the universe.
Lattice QCD results have been incorporated into an equation of state%
~\cite{Laine:2006cp} which is  used to compute sterile 
neutrino production~\cite{Asaka:2006nq}. 
The sterile-neutrino reaction rate depends both on the real 
and the imaginary part of the  self-energy of the {\it active} neutrinos.
The latter receives contributions from interactions with leptons and
with quarks.
The quark contribution is determined by susceptibilities of 
the QCD plasma's conserved charges.
Lattice determinations of susceptibilities 
\cite{Bazavov:2012jq, Borsanyi:2011sw} have been included into 
sterile-neutrino evolution equations in \cite{Venumadhav:2015pla}. 
The most difficult to compute are hadronic contributions to the imaginary 
part of the active-neutrino self-energy, or opacity, 
which can be written  as a momentum integral over mesonic spectral functions 
\cite{Asaka:2006rw}. 
Determining these spectral functions on the lattice is challenging.
Previous works on resonant sterile-neutrino production 
\cite{Venumadhav:2015pla,Ghiglieri:2015jua} have treated the opacity in 
different approximations, but it still remains an open question how important 
the non-perturbative contributions are. 
In this work we address it by computing the sterile-neutrino
dark matter production for various lepton asymmetries and by carefully
treating the resonances which appear during the time evolution.
 
After introducing our setup in sec.~\ref{s:setup}, we state in 
sec.~\ref{s:eom} evolution equations for the sterile-neutrino phase space 
densities and lepton asymmetries.
We will include various leptonic and hadronic active-neutrino self-energy 
contributions in sec.~\ref{s:active}. 
We solve the coupled system of equations for various values of initial 
lepton asymmetries in sec.~\ref{s:num}. 
Motivated by our findings we update the lower limit on the active-sterile 
mixing angle in the two-flavor (one active and one sterile flavor) 
scenario by using maximal values of lepton asymmetries allowed by BBN. 
Conclusions are in sec.~\ref{s:concl}. 
Appendix \ref{a} contains additional figures showing how  
the resonances emerge when the lepton asymmetry is increased. 

\section{Non-equilibrium evolution equations}	\label{s:evo}

\subsection{Setup}	\label{s:setup}

We consider the Standard Model augmented by one family of sterile Majorana
neutrinos $N$ with Majorana mass $M$ and 
non-zero Yukawa couplings $h_\alpha$ to all active neutrino flavors,
\begin{equation}
  \label{Lagrange} 
   \mathcal{L} 
   = 
	\mathcal{L}_{SM} + \frac{1}{2} \bar{N} (i \slashed \partial - M) 
   N 
   - \sum_\alpha 
   \left (
      \bar{N} 
   \widetilde{\varphi}^{\dagger} h_\alpha \ell_\alpha
    + { \rm H.c.} \right ) 
\end{equation}
where $\widetilde{\varphi} = i \sigma^2 \varphi^*$ is the 
conjugate Higgs doublet 
and $\ell_\alpha = (\nu _{_L{}_\alpha}, e_{_L{}_\alpha})^\top$ the 
left-handed lepton doublet.
The sterile-neutrino field in the interaction picture reads 
\begin{equation}
  N(x)
  = \sum_{\vec k, \lambda} \frac{1}{\sqrt{2 k^0 V}}
    \Big[ e^{-i k \cdot x}
      u^{}_{\vec k \lambda} a^{}_{\vec k \lambda} + e^{i k \cdot x} v^{}_{\vec k \lambda} a^\dagger_{\vec k \lambda}
      \Big]
    ,
   \label{N} 
\end{equation}
with the energy $k^0 = ( \vec k^2 + M^2) ^{ 1/2 } $.  The spinors $ u $
and $ v $ satisfy the Majorana condition $u = v^c $, where $ c $ 
denotes charge conjugation. 
Furthermore, the creation-/annihilation operators fulfill 
$\{a^{}_{\vec k \lambda}, a^\dagger_{\vec q \lambda'}\} 
= \delta_{\vec k, \vec q} \delta_{\lambda \lambda'}$ with helicities 
$\lambda = \pm 1/2$. 
$V$ denotes the volume of our system. 
With these operators we define the sterile-neutrino
phase space density operators as
\begin{equation}
   f^{}_{\vec k \lambda} \equiv a^\dagger_{\vec k \lambda  } a^{}_{\vec k \lambda  }.
\end{equation}
The number operator for left-handed leptons of flavor $\alpha$ reads
\begin{equation}
   L_\alpha = \int d^3x \;\ell^\dagger_\alpha \ell^{}_\alpha
   .
   \label{L} 
\end{equation}
We will start the time evolution  at temperatures of a few GeV, 
where sphaleron processes have long terminated. 
Then baryon number $B$ is conserved, and its tiny value can be 
well approximated by zero for our purposes. 
Furthermore electric charge $Q$ is conserved and exactly zero.

\subsection{Equations of motion}	\label{s:eom}

The expectation values of the operators introduced above 
are conserved by Standard Model interactions and thus evolve much 
more slowly  than the other degrees of freedom. 
Their deviations from equilibrium characterize the non-equilibrium
state, and are assumed to be small.
We introduce chemical potentials
$\mu^{}_{L_\alpha}$, $\mu^{}_B$ and $\mu_Q$ and denote them  collectively
by $\mu$ in the following.
In the infinite volume limit the phase space densities and the lepton number densities 
$n _ { L _ \alpha  } \equiv L_\alpha /V$ satisfy the evolution 
equations \cite{Ghiglieri:2019kbw,Bodeker:2019rvr}
\begin{align} 
   \dot{f}_{\vec k \lambda  }
    = - 
   \frac{1}{2 k^0}
   \sum_\alpha 
   &\Big\{  
   \bar{u}_{\vec k \lambda}\rho_{\alpha }(k, \mu)u_{\vec k \lambda}
    \big [ 
       f_{\vec k \lambda  } - f _ { \rm F } ( k^0 - \mu  _ { L _ \alpha } )
    \big ] 
   \nonumber \\
   & {}
   + 
\bar{v}_{\vec k \lambda}
   \rho_{\alpha}(-k, \mu)v_{\vec k \lambda}
    \big [ 
       f_{\vec k \lambda  } - f _ { \rm F } ( k^0 + \mu  _ { L _ \alpha } )
    \big ]
    \Big \}
    ,
   \label{fdot}
\end{align}
and
\begin{align}
   \dot n _ { L _ \alpha  } 
   =
   \sum _ \lambda  \int \! \frac { d ^ 3 k } { ( 2 \pi  ) ^ 3 2 k^0} 
   &\Big \{ 
   \bar{u}_{\vec k \lambda}\rho_{\alpha }(k, \mu)u_{\vec k \lambda}
      \big [ f _ { \vec k \lambda  } 
         - f _ { \rm F } ( k^0 - \mu  _ { L _ \alpha } )
    \big ] 
   \nonumber \\
   &   { }
   -   
   \bar{v}_{\vec k \lambda} \rho_{\alpha}(-k, \mu)v_{\vec k \lambda}
    \big [ 
       f_{\vec k \lambda  } - f _ { \rm F } ( k^0 + \mu  _ { L _ \alpha } )
    \big ] 
   \Big \} 
	,
   \label{ndot}
\end{align}
with the Fermi-Dirac distribution $f _ {\rm F}(x) = 1/(e^{x/T}+1)$
and the spinors $ u $ and $ v $ which appear in (\ref{N}).
Furthermore,
\begin{align}
     \rho _ \alpha ( k, \mu ) 
     \equiv
     \frac 1 { i } 
    \left [ 
   \Delta ^ { \rm ret } _ { \alpha } ( k , \mu ) -
     \Delta  ^ { \rm adv } _ { \alpha } ( k , \mu )
   \right ] 
   \label{rhoal}
   ,
\end{align} 
is the spectral function which is determined by the retarded and advanced  
2-point function
\begin{equation}
\Delta^{\rm ret, adv}_{\alpha  }(k ) 
   = 
   \pm i \int d ^ 4 x \, \Theta  ( \pm t ) 
   e ^{  i k x } 
   \Big \langle  \Big \{  J _ \alpha  ( x ) , \bar J _ \alpha  ( 0 ) 
      \Big\}  \Big \rangle
   \label{retadv} 
\end{equation}
of the operator $ J _ \alpha  \equiv
\widetilde{\varphi}^{\dagger} h_\alpha \ell_\alpha $, which couples to $ \bar N $ in (\ref{Lagrange}).

Oftentimes the kinetic equations are expanded in $ \mu  $. 
However, for a proper treatment of resonances, which show up in the spectral 
functions $\rho_{\alpha }$, 
one has to include all orders in $ \mu  $,
which will become more apparent below. 
The relation of the chemical potentials to the charge densities,
to be discussed 
in sec.~\ref{s:active}, can still be assumed to be linear. 

In the broken phase, where $\avg{\widetilde{\varphi}}=(\varv/\sqrt{2},0)^\top$ 
with $\varv = 246$ GeV, the 2-point function 
(\ref{retadv}) is proportional to the active-neutrino propagator,
\begin{equation}	
   \label{nuprop}
   \Delta_  \alpha 
   ( k, \mu ) 
   =
   \theta_\alpha^2 M^2 P_{ \rm L }
  \frac{-1}{ \slashed k - \Sigma_\alpha ( k , \mu)} 
   P_ { \rm R } 
   .
\end{equation}
Here we have factored the chiral projectors 
$P_{\rm R,L} = \frac{1}{2} ( 1 \pm \gamma_5 )$ out of the propagator 
and furthermore introduced the active-sterile mixing angle
\begin{equation}
\theta_\alpha \equiv \frac{\abs{h_\alpha} \varv}{\sqrt{2} M}
   \label{theta} 
   .
\end{equation}
The active-neutrino self-energy in the plasma rest frame can be
approximated as%
\cite{Weldon:1982bn,Ghiglieri:2015jua}
\begin{equation}
    \label{Sig} 
    \Sigma_\alpha ^ { \rm ret} ( \pm k, \mu )    
   = 
   \gamma^0 \bigg( \mp b_\alpha + c_\alpha 
     - \frac{i \Gamma_\alpha}{2} \bigg) 
\end{equation}
with real $ b _ \alpha $, $ c _ \alpha $.
$ \Gamma _ \alpha $~is 
the imaginary part of the refractive index for the active neutrinos 
\cite{Notzold:1987ik}, and is also referred to as neutrino opacity.
$ \Gamma  _ \alpha  /2 $ is the neutrino damping rate 
(see e.g.~\cite{Bellac:2011kqa}). 
The function $ c _ \alpha $ is odd in $ \mu  $. 
For the chemical potentials under consideration we only 
need to keep the linear
order in $ \mu  $ for $ c _ \alpha $, and we can neglect 
the $ \mu  $-dependence of $ b _ \alpha  $ and $ \Gamma  _ \alpha  $. 
The advanced self-energy is obtained from~(\ref{Sig}) by replacing
$ i \Gamma  _ \alpha  \to -i \Gamma  _ \alpha  $.
Then one obtains the spectral function
\begin{equation}
   \label{broken}
   \rho_ \alpha   ( \pm k, \mu)
   =
   \frac{ \theta_\alpha^2 \Gamma_\alpha M^2 
      }
        { [ M^2 + 2 k^0 ( b_\alpha \mp c_\alpha ) ]^2 
         + (k^0 \Gamma_\alpha)^2} 
   P_{\rm L}
   ( 2 k^0 \slashed k  - M^2 \gamma^0 )P_{ \rm R} 
   .
\end{equation}
A quick calculation yields
\begin{equation}	
   \label{ubru}
\bar{u}_{\vec k \pm} \rho_{\alpha }(k, \mu)u_{\vec k \pm} 
  = 
   \frac{\theta_\alpha^2 \Gamma_\alpha M^4 ( k^0 \mp | \vec k | )}
   {[M^2+2 k^0 (b_\alpha-c_\alpha) ]^2 + ( k^0 \Gamma_\alpha)^2}
   ,
\end{equation}
\begin{equation} 
   \bar{v}_{\vec k \pm} \rho_ \alpha  (-k, \mu)v_{\vec k \pm} 
    = 
   \frac{\theta_\alpha^2 \Gamma_\alpha M^4 ( k^0 \pm | \vec k | ) }
   {[M^2+2 k^0 (b_\alpha+c_\alpha)]^2 + ( k^0 \Gamma_\alpha)^2}
   .
   \label{vbrv}
\end{equation} 
Resonances occur when a square bracket in the  
denominator of (\ref{ubru}) or (\ref{vbrv})  vanishes. 
This can happen when   $c_\alpha$
is large enough and has the appropriate sign. 
For a given sign of an initial lepton asymmetry, only one of the expressions
\eqref{ubru} and \eqref{vbrv} can lead to resonances.
Moreover, at leading order in $M/| \vec k | $ 
only the terms containing
$\bar{u}_{\vec k -} \rho_{\alpha}u_{\vec k -}$  
or $\bar{v}_{\vec k +} \rho_{\alpha}v_{\vec k +}$
will contribute in the evolution equations. The subleading terms will
be dropped in the following.  

Due to the isotropy of the universe
the phase space density only depends  on $ | \vec k | $.
Then the  Hubble expansion is taken into account by replacing
\begin{equation}
   \dot{f}_{\vec k \lambda} \rightarrow \Big( \partial_t - H | \vec k |
     \partial_{ | \vec k | }  \Big) f_{ \vec k \lambda} 
   .
   \label{fdotH}
\end{equation}
The Hubble parameter is given by
\begin{equation}
   H = \sqrt{ \frac{ 8 \pi \rho }{ 3 M_{\rm Pl}^2 } }
   \label{H} 
   ,
\end{equation} 
where $\rho $ is the energy density and $M_{\rm Pl} \simeq 1.22 \cdot 10^{19}$~GeV is the Planck mass.
By using $ K \equiv | \vec k | a ( t ) / a (t_{\rm end}) $ as an
independent variable, 
where $ a $ is the scale factor, (\ref{fdotH}) 
turns into $ \partial _ t f _{K \lambda} $ and the equation for 
$ f $ becomes an ordinary differential equation. 
$a (t_{\rm end})$ is 
the scale factor at the time corresponding to the 
temperature $ T_{\rm end} = 10$ MeV at which we compute the final
abundances. 
For lepton number densities the Hubble expansion is taken into account through
the replacement
\begin{equation}
\dot{n}_{L_\alpha} \rightarrow \Big( \partial_t + 3 H \Big) n_{L_\alpha}
   .
   \label{ndotH} 
\end{equation}
The term proportional to $ H $ is eliminated by considering the
differential equation for  $n_{L_\alpha}/s$ where $s$ is the entropy density.
Finally, the 
time derivatives  are replaced by  temperature derivatives
via $dT/dt=-TH(T)3c_s^2(T)$,  with the speed of sound  $c_s$.

\subsection{Active-neutrino self-energy}	
   \label{s:active}

For vanishing chemical potentials 
the real part of the active-neutrino self-energy arises at 
$\mathcal{O}\big( G _ { F}/m_W^2 \big)$, where $ G _ { F } $
is the Fermi constant and $ m _ W $ is the $ W $-boson mass  
\cite{Notzold:1987ik, Asaka:2006nq},
\begin{equation}
   \label{b} 
b_\alpha = \frac{ 8 \sqrt{2}  G _ { F} }{m_W^2} k^0 
   \bigg[ \cos^2 \theta_{\rm W} \frac{7 \pi^2 T^4}{360} 
    + \! \int \! \frac{d^3 p}{(2 \pi)^3} \frac{f _ {\rm F} (
  E_ { \alpha })}
   {E_ { \alpha }} 
   \bigg( \frac{4}{3} \vec p^2 + m_ { \alpha }^2 \bigg) \bigg],
\end{equation}
where $ \theta  _ { \rm W } $ is the weak mixing angle, 
$ E_ { \alpha } 
\equiv  ( \vec p^2 + m_ { \alpha }^2 ) ^{ 1/2 } $, 
and $m_ { \alpha }$ is the mass of the charged lepton of flavor 
$\alpha$.
We have neglected the masses of active neutrinos.
$ b _ \alpha  $ is positive, which corresponds to an index of refraction
greater than 1, or a negative thermal mass squared.

The leading contribution due to non-zero chemical potentials 
can have either sign. It arises at $\mathcal{O}\big( G _ {  F }  \big)$
\cite{Notzold:1987ik, Venumadhav:2015pla, Ghiglieri:2015jua},
\begin{align}	
c_\alpha 
   =  
   \sqrt{2} G _ {F} 
   \bigg[ &
   2 n _{\nu  _\alpha} 
   + \sum_{\beta \neq \alpha} n _{\nu _ \beta} 
   + \bigg( \frac{1}{2} + 2 \sin^2 \theta_{\rm W} \bigg) n _ { e _ \alpha }
   - \bigg( \frac{1}{2} - 2 \sin^2 \theta_{\rm W} \bigg) 
      \sum_{\beta \neq \alpha} n _ { e _ \beta } 
  \nonumber
   \\ 
 &
  -\frac{1}{2} n_B 
  + \bigg( 1 - 2 \sin^2 \theta_{\rm W} \bigg) n_Q^ { \rm had }  \bigg]
   ,
   \label{cn}
\end{align}
without the $ 1/m ^ 2 _ W $ suppression of (\ref{b}). 
Therefore it can be of similar size as (\ref{b}) when the chemical
potentials are small.
$ n  _ { \nu   _ \alpha  } $, 
$ n  _ { e _ \alpha  } $ are particle minus anti-particle
number densities of neutrinos and charged leptons.
They can be written in terms of the particle
chemical potentials $ \mu  _ i $, 
\begin{equation}	
   n _ i =  \chi _ i \mu_ i 
   \label{na}
   ,
\end{equation}
where the lepton susceptibilities $ \chi _ i $  can be evaluated in the ideal 
gas limit, 
\begin{align}
\chi^{ } _ { e_ \alpha } &= -2 g _ { e _ \alpha } \int \frac{d^3 p}{(2 \pi)^3} 
  f'_ { \rm F }  (E_ { \alpha }),
	\label{chie}  \\
\chi^{ } _ { \nu _ \alpha  } &= g _ { \nu _ \alpha  } \frac{T^2}{6},
	\label{chinu} 
\end{align}
with $ g _ { \nu _ \alpha  } = 1$, $ g _ { e_ \alpha } = 2$.
The hadronic contribution to the electric charge density $n_Q^ { \rm had } $ 
can be written as
\begin{align} 
   n _ Q ^ { \rm had }  = \chi  _ { QQ } ^ { \rm had }  \mu ^ { }  _ Q + \chi ^{} _ { QB } \mu ^ { }  _ B 
   ,
\end{align} 
where $ \chi  _ { QQ } ^ { \rm had }  $ is the hadronic contribution to the
electric-charge susceptibility.
The particle  chemical potentials in \eqref{na} 
can be written in terms of the chemical potentials of the slowly
varying and of the conserved charges,
\begin{align}
\mu^{}_ { e_ \alpha } &= \mu ^ { } _{L_{\alpha}} - \mu^{}_Q, 
  \label{mue} 
   \\
\mu^{}_{\nu_{\alpha}} &= \mu ^ { } _{L_{\alpha}}.
  \label{munu} 
\end{align}
The latter can be expressed through  the lepton number densities
$ n _ { L _ \alpha  } $
by inverting 
\begin{align}
  n_B &= \chi ^ {} _{BB} \mu^{}_B +\chi^ {} _{QB} \mu^{}_Q 
   ,	
   \label{n_B} \\	
  n_Q &=  
   \left (\chi _{QQ}^{\rm had}+\chi_{QQ}^{\rm lep}\right ) \mu^{}_Q
   +\chi^{ } _{QB}\mu^{}_B
   +\chi^{}_{QL_\alpha}\mu_{L_\alpha} 
   ,
	\label{n_Q} \\
n_{L_\alpha} 
   &= 
   \chi^{}_{L_\alpha L_\alpha}\mu ^ { } _{L_\alpha}
   + \chi^{}_{Q L_\alpha}\mu^{}_Q
   ,	
   \label{n_L}
\end{align}
and assuming 
vanishing overall baryon and electric charge density, $ n _ B = n _ Q = 0 $.
The leptonic part of the electric charge susceptibility in 
(\ref{n_Q}) can be written
in terms of (\ref{chie}), 
\begin{equation}
\chi_{QQ}^{\rm lep } = \sum_{\alpha = e,\mu,\tau} \chi ^ { } _ { e_ \alpha }
    .
   \label{chilep} 
\end{equation}
The susceptibilities in (\ref{n_L}) are related to (\ref{chie}),(\ref{chinu})
by
\begin{align}
\chi ^ { } _ {L_\alpha L_\alpha} 
   & = 
   \chi ^ { } _ { e_ \alpha } + \chi ^ { } _{\nu_\alpha}
   \label{chiLL} 
    ,\\
   \chi ^ { } _{Q L_\alpha} 
   & = 
    - \chi ^ { } _ { e_ \alpha } 
   \label{chiQL} 
    .
\end{align}
The susceptibilities $
\chi_{QQ}^{\rm had}$, $ \chi ^ { } _{BQ}$, 
and $ \chi ^ { } _{BB}$ have been determined 
on the lattice for temperatures near the QCD
crossover \cite{Bazavov:2012jq, Borsanyi:2011sw}. 
Reference \cite{Venumadhav:2015pla} has used a hadron resonance gas model below and perturbation theory  above and connected all three regions via spline interpolations, which we are going to use.%
\footnote{Available 
at \url{https://github.com/ntveem/sterile-dm/tree/master/data/tables}.}

The dominant contribution to the active-neutrino opacity
$\Gamma_\alpha$ appears at 
$\mathcal{O}\big( G_F^2 \big)$ since the $\mathcal{O}\big( G_F \big)$ 
contributions are suppressed by $ \exp ( -m _ W/T ) $.
It can be split into a leptonic and a 
hadronic piece, 
\begin{align}
   \label{Gsum} 
   \Gamma^{}_\alpha 
	= 
   \Gamma_\alpha^{\rm lep} + \Gamma_\alpha^{\rm had} 
   .
\end{align} 
Over a large part of the 
temperature range which is relevant for sterile-neutrino production, 
$\Gamma_\alpha^{\rm had}$ is non-perturbative. 
Two different approaches have been taken to calculate this function. 
In \cite{Asaka:2006nq} the free-quark approximation 
is used for the whole temperature range,
but, in order to account for the strong interaction, 
the number of colors $ N _ c $ is replaced by a  temperature dependent
$ N_{c,\rm eff} ( T )  $ which vanishes at low temperatures,  
and equals 3 at the highest  temperature. 
In \cite{Venumadhav:2015pla}, on 
the other hand, the free-quark approximation at high temperatures is connected 
to chiral perturbation theory at low temperatures via spline interpolations. 
We refer to these two approximations for (\ref{Gsum})  as 
$\Gamma^{N _ { c ,\rm eff}}$ and $\Gamma^{\rm \spline}$. 
We will also consider the approximation $ \Gamma _ \alpha   ^{ \rm had } =0 $
for which we write $ \Gamma  ^ { \rm lep } $.

\section{Numerical results}	\label{s:num} 

We integrate the coupled eqs.~\eqref{fdot}, \eqref{ndot} 
from $ T = 4$ GeV down to 10 MeV, 
using the  parameterizations of the energy $ \rho (T) $  
and entropy density 
$ s (T) $ 
as well as the speed of sound $c_s(T)$ from \cite{Asaka:2006nq},%
\footnote{Available at 
\url{www.laine.itp.unibe.ch/dmpheno/release_2016jun21.tar.gz}. 
\label{footnote_Laine}} 
based on calculations in \cite{Laine:2006cp}. 

Depending on the value of the lepton asymmetry, high numerical precision is
needed to handle sharp resonances. 
In practice, the numerical solution requires an implicit multi-step 
method to deal with possible equation stiffness. 
We assume that at the starting temperature there are no sterile 
neutrinos, i.e., $f_{\vec k \lambda  } = 0$.
We recast the mixing angles into the total active-sterile mixing angle through
\begin{equation}
   \label{sinth} 
\sin^2 (2\theta) \approx 4 \sum_\alpha \theta^2_\alpha,
\end{equation}
which is the quantity that one can constrain experimentally from X-ray 
observations. 

\subsection{Influence of the opacity on sterile-neutrino production
		}
    \label{s:width}
In this section we make the simplifying assumption that only one Yukawa 
coupling is non-zero, namely $h_{\mu}$. 
This allows us to compare the effect of the opacities  
$\Gamma_\alpha^{ N_{ c,\rm eff}}$, which are available for 3 lepton flavors, 
and 
$\Gamma_\alpha^{\rm spline}$, which 
is currently only available for $\alpha = \mu$.
Then only the term with $ \alpha  = \mu  $ contributes in \eqref{fdot}, 
and only the muon-flavor asymmetry will be dynamical. 
In principle, non-zero electron and tau-flavor asymmetries 
can influence the evolution equations as they appear in the 
functions $c_\alpha$, but we assume these to be zero.
We choose $M = 7.1$ keV, motivated 
by the tentative signal reported in~\cite{Bulbul:2014sua,Boyarsky:2014jta},
and $\theta_\mu^2 = 2.5 \cdot 10^{-13}$ as a 
representative point in the available parameter space.

\begin{figure}[t]
\begin{minipage}{.49\textwidth}
	\centering
	\includegraphics[width=\textwidth]{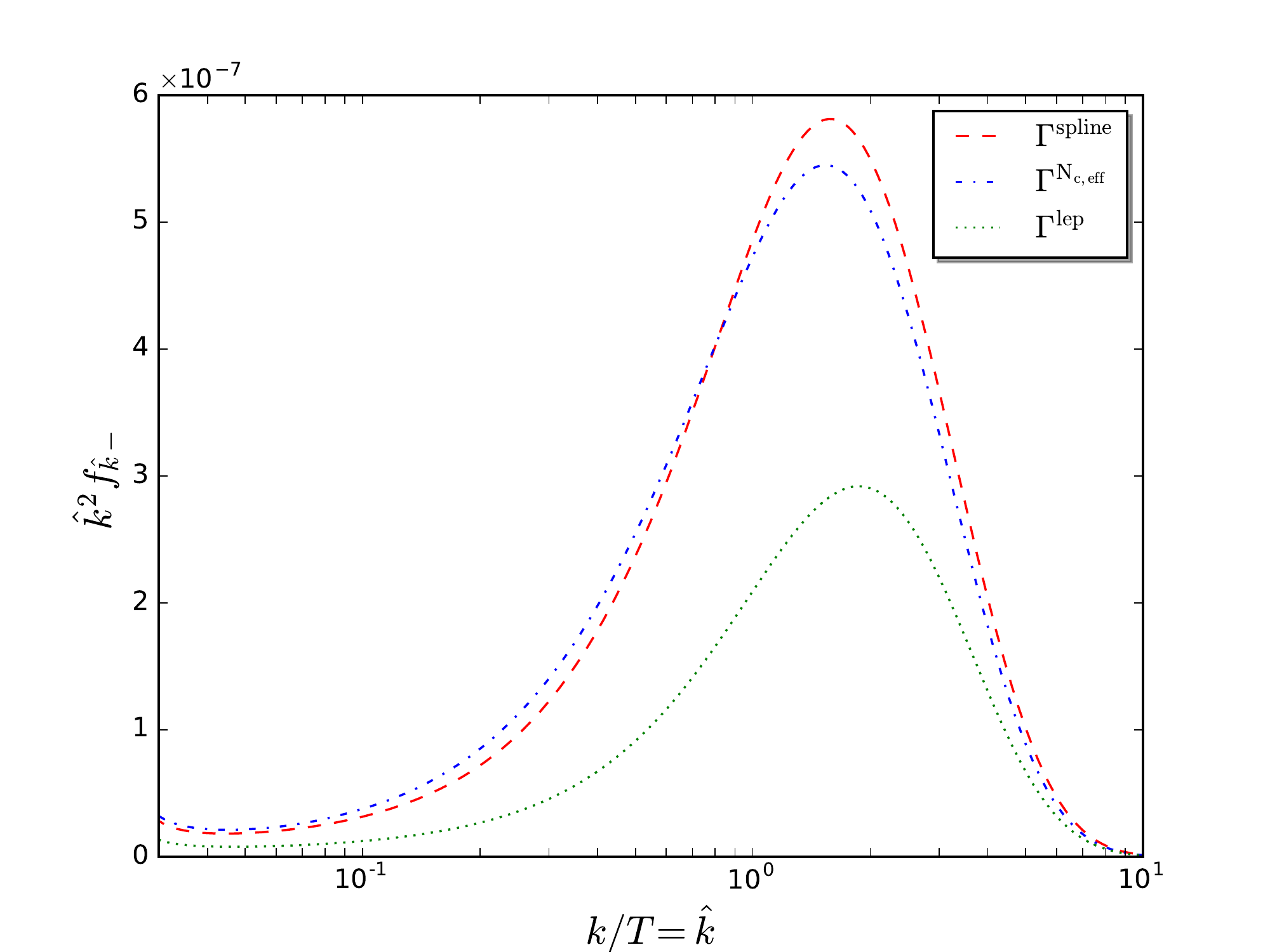}
\end{minipage}
\begin{minipage}{.49\textwidth}
	\centering
	\includegraphics[width=\textwidth]{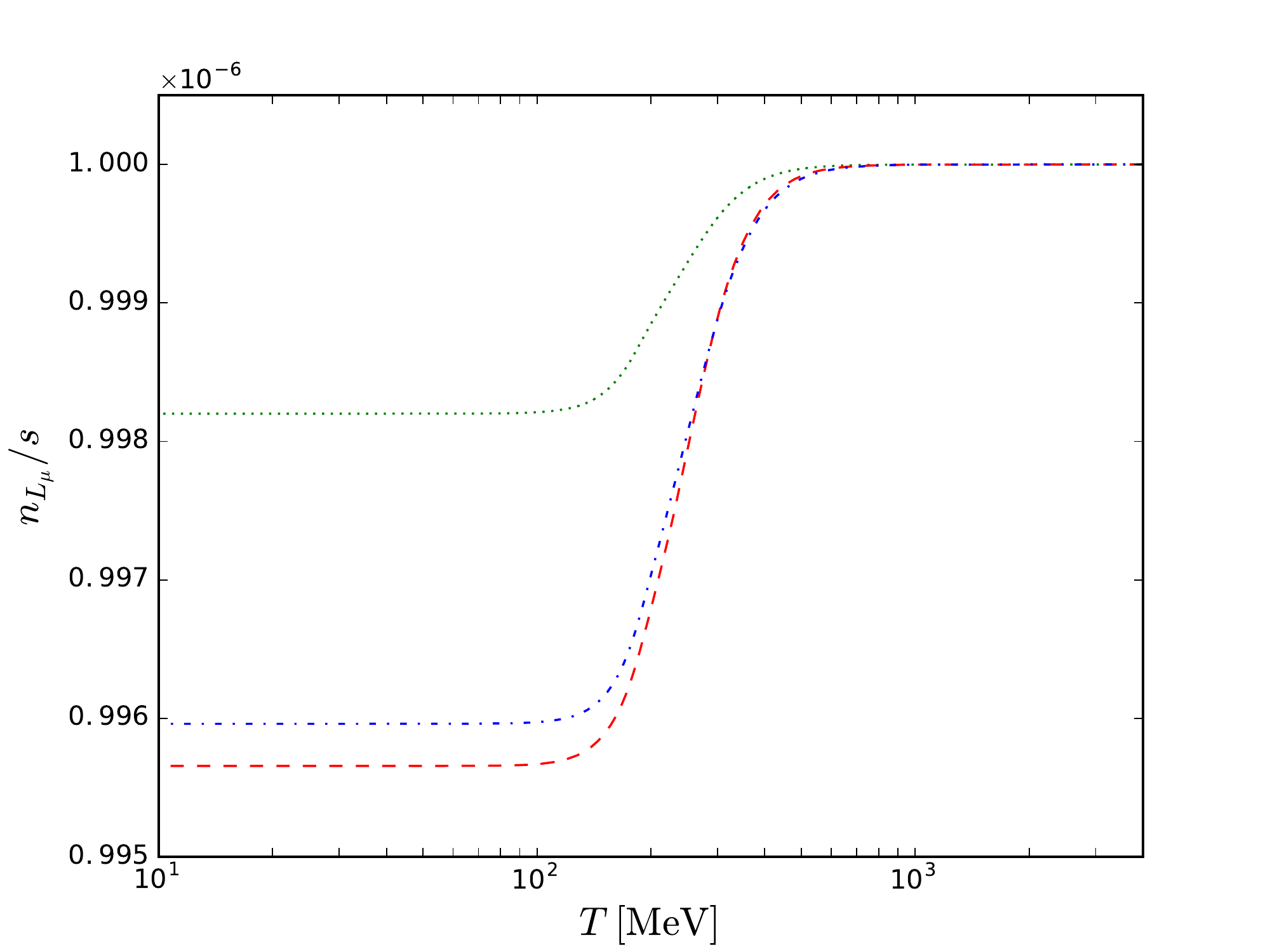}
\end{minipage}
	\caption{Solutions to the kinetic equations for $M = 7.1$keV, 
   $\theta_\mu^2 = 2.5 \cdot 10^{-13}$. The different
   curves are obtained with the different approximations for 
   $ \Gamma  _ \alpha  $. 
   Left: Phase space densities of sterile neutrinos
with negative helicity at $T = 10$ MeV. 
The fraction of the dark matter energy density is 
$(\Omega_{ \rm s} /\Omega_{\rm DM})_{\rm spline} = 5.7 \cdot 10^{-4}$, 
$(\Omega_{ \rm s} /\Omega_{\rm DM})_{ N_{ c,\rm eff}} = 5.3 \cdot 10^{-4}$, 
$(\Omega_{ \rm s} /\Omega_{\rm DM})_{\rm lep} = 3.2 \cdot 10^{-4}$ 
respectively. 
Right: Evolution of the lepton asymmetry. }
\label{f:sol1e6}
\end{figure}

We compare the resulting sterile-neutrino phase space densities at 
$T = 10$ MeV obtained with the different opacities 
$\Gamma_\mu^{\rm spline}, \Gamma_\mu^{ N_{ c,\rm eff}}$, and 
$\Gamma_\mu^{\rm lep}$ for a set of different positive 
initial values for $n_{L_\mu}/s$. 
The opacities
$\Gamma_\mu^{\rm spline}$ and $\Gamma_\mu^{\rm lep}$ are available for 
momenta $10^{-4} \leq |\vec k|/T \leq 20$, whereas $\Gamma_\mu^{N_{c,\rm eff}}$ is available for $0.03 \leq |\vec k|/T \leq 12.5$. 
We use the latter range when solving the evolution equations, which 
is sufficient for our purposes.
For positive lepton asymmetries, resonances mainly contribute to the 
production of 
sterile neutrinos with negative helicity, 
while for positive helicity the resonant contribution is suppressed 
with $ M/| \vec k | $. 
If there are resonances, then there are usually two resonance
frequencies for each $ | \vec k | $
\cite{Abazajian:2001nj,Ghiglieri:2015jua,Venumadhav:2015pla}. 
For most of the relevant temperatures, 
the two resonances lie in the momentum range we consider. 
In practice, the smaller resonance frequency dominates the sterile-neutrino 
dark matter production, and the larger
one plays a negligible role~\cite{Venumadhav:2015pla}.

We show results for three different initial values of $n_{L_\mu}/s $
in figs.~\ref{f:sol1e6}, \ref{f:sol1e5} and \ref{f:sol3e4},
additional ones can be found in appendix \ref{a}.
Generally we observe that the higher the initial lepton asymmetry, 
the larger the phase space densities become.
In fig.~\ref{f:sol1e6} the lepton asymmetry is so low that resonances 
are outside the  displayed momentum range
(the dominant one leads to the slight increase at small momenta) 
and only give a small contribution to the production.
In contrast, the initial lepton asymmetry in fig.~\ref{f:sol1e5} is high 
enough so
that each momentum mode in the shown range passes through a resonance, 
giving much larger phase space densities. 
The same is true for fig.~\ref{f:sol3e4}, 
where we chose the initial asymmetry such that we obtain the 
complete dark matter abundance.
In fig.~\ref{f:sol1e6} we see how the different approximations 
for the opacity influence the
sterile-neutrino production and in parallel the depletion of $n_{L_\mu}/s$. 
The purely leptonic contribution is the smallest one, 
resulting in the least efficient production.
In figs.~\ref{f:sol1e5} and \ref{f:sol3e4} one can 
see that for a larger initial 
lepton asymmetry,
there is only a sub-percent difference in the final abundance of 
sterile neutrinos between using the full opacity and using 
only the leptonic contribution. 
The resulting phase space densities have become indistinguishable. 

\begin{figure}[t]
\begin{minipage}{.49\textwidth}
	\centering
	\includegraphics[width=\textwidth]{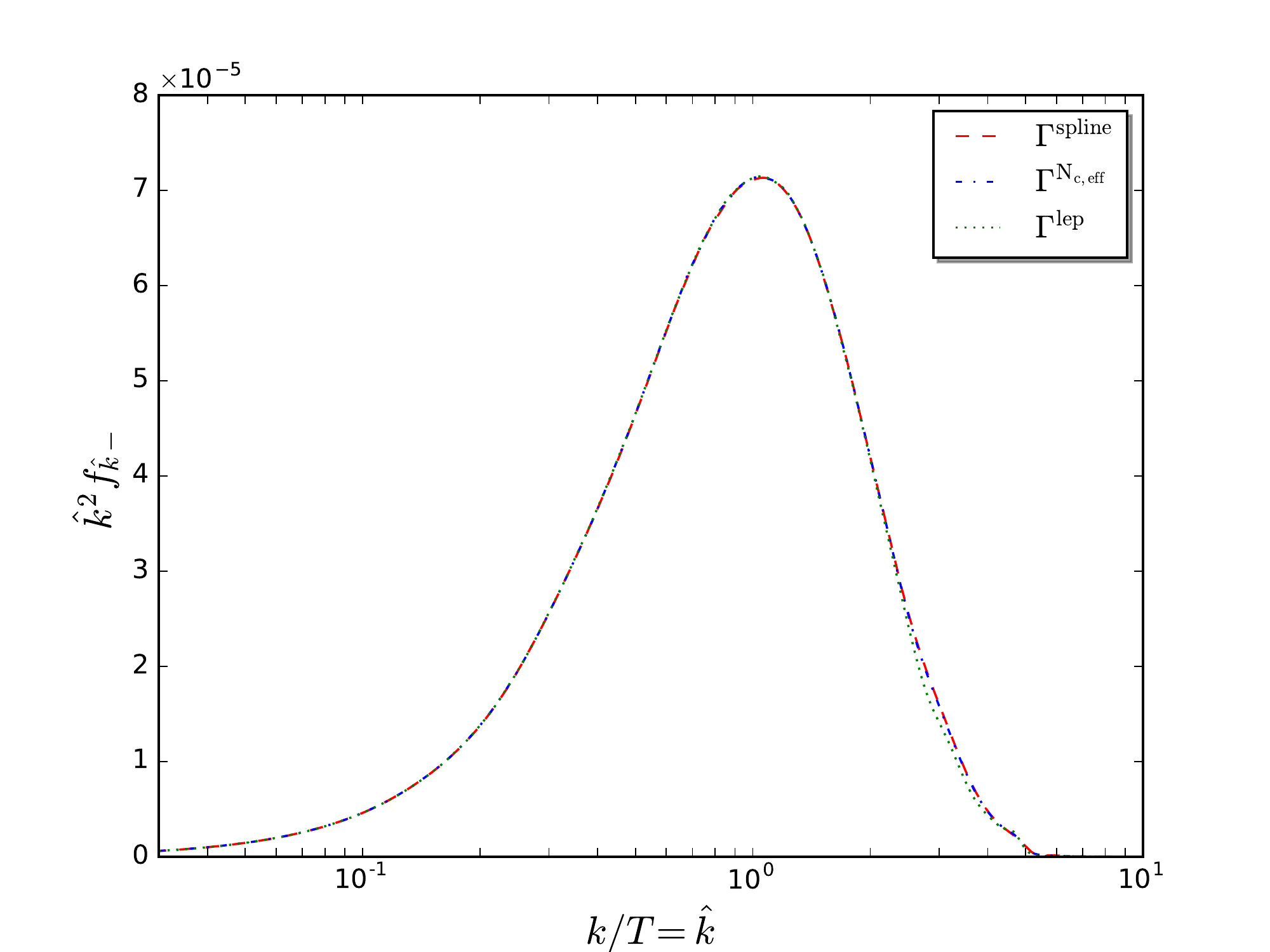}
\end{minipage}
\begin{minipage}{.49\textwidth}
	\centering
	\includegraphics[width=\textwidth]{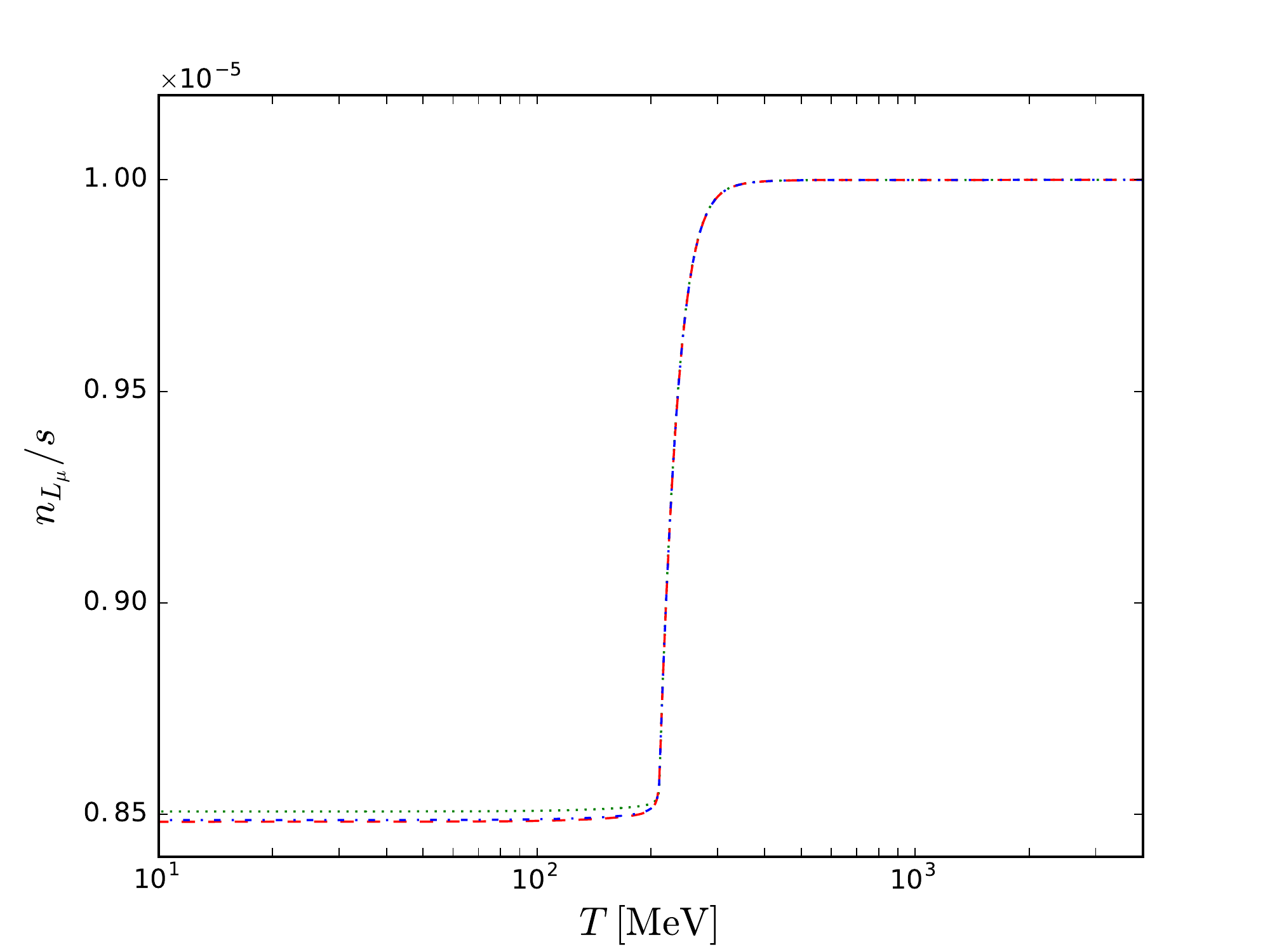}
\end{minipage}
	\caption{Same as fig.~\ref{f:sol1e6} but with a higher initial 
   lepton asymmetry. The fraction of the dark matter energy density in all three cases is roughly $\Omega_{ \rm s} /\Omega_{\rm DM} = 2.5 \cdot 10^{-2}$.}
\label{f:sol1e5}
\end{figure}

In the limit $ \Gamma  _ \alpha \to 0 $
\eqref{ubru} turns into a delta function
\cite{Ghiglieri:2015jua}. 
This indicates that  the dominant 
resonance in figs.~\ref{f:sol1e5} and \ref{f:sol3e4} is so sharply peaked, that 
the differences in the active-neutrino opacities become irrelevant. 
The same is true for the lepton asymmetry evolution. 
No matter what opacity is used, the depletion is almost identical. 
We find that this behavior occurs in all of the 
allowed (white) parameter space shown in 
fig.~\ref{f:bounds},
if we tune the lepton asymmetry such that the resulting sterile neutrino 
energy density gives the correct dark matter abundance. 
The differences in energy densities obtained with the different 
opacities
are typically below the 2\%-level, for very low masses and high 
mixing angles at most 5\%.
The transition from quite different to basically equivalent solutions 
by increasing the lepton asymmetry can be followed 
in smaller steps in appendix \ref{a}.

\begin{figure}[t]
\begin{minipage}{.49\textwidth}
	\centering
	\includegraphics[width=\textwidth]{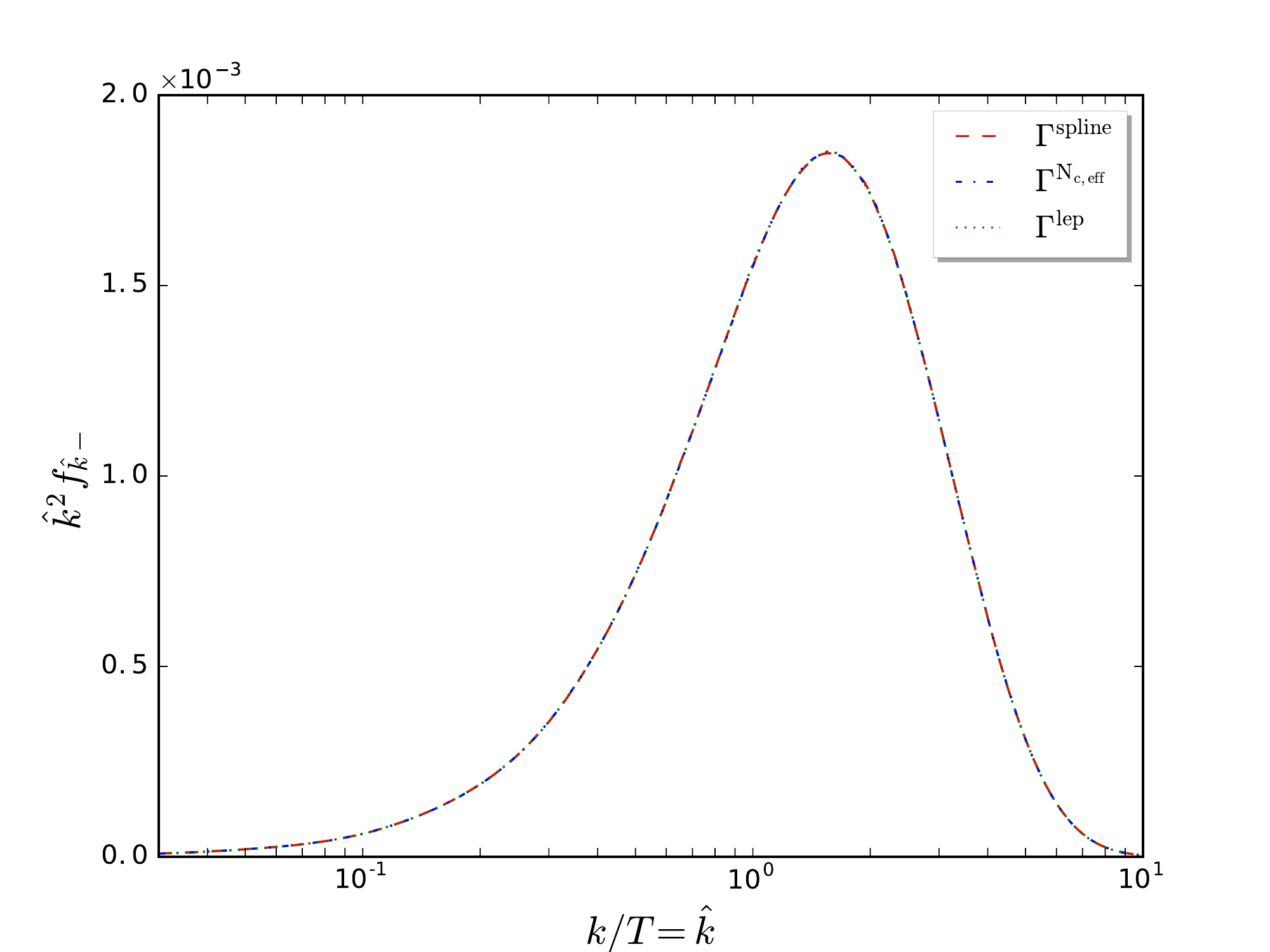}
\end{minipage}
\begin{minipage}{.49\textwidth}
	\centering
	\includegraphics[width=\textwidth]{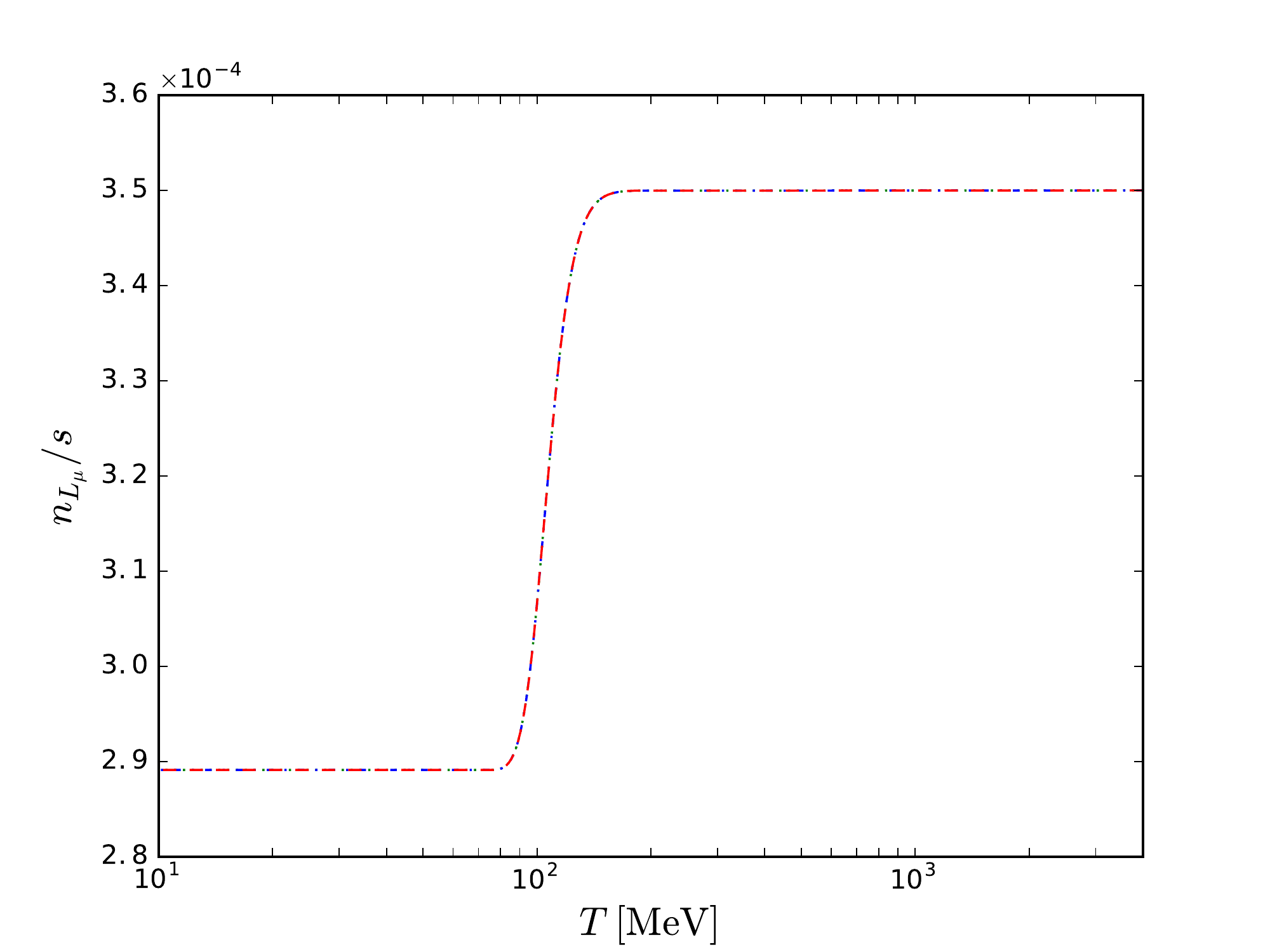}
\end{minipage}
	\caption{Same as fig.~\ref{f:sol1e6} but with an initial 
   lepton asymmetry tuned such that the sterile neutrino energy density gives the complete relic dark matter abundance, $\Omega_{ \rm s} = \Omega_{\rm DM}$.}
\label{f:sol3e4}
\end{figure}

We have used the publicly available code of \cite{Ghiglieri:2015jua} to check 
our calculation, and what we find is 
mostly in agreement with our results described above. 
For very high asymmetries we find that the resulting 
phase space densities suffer from sporadic kinks, hinting at 
numerical instabilities which we could not get rid of by naively 
increasing the desired precision. 
Nevertheless the resulting figures resemble ours quite well. 

Our findings partly disagree with the ones in \cite{Venumadhav:2015pla}, 
which were
calculated using {\it sterile-dm}, a publicly available code created by 
the authors 
of~\cite{Venumadhav:2015pla}. 
It uses 1,000 momentum bins as a default, which apparently misses parts 
of the resonances in the sterile-neutrino production. 
While this problem is absent for non-resonant production, it
becomes more and more severe for increasing asymmetry. 
We have explicitly checked that increasing the number of momentum bins 
to 30,000 gives results which mainly agree with ours.
This problem could be the cause of the rather large 
differences in the phase space densities using either 
$\Gamma_\alpha^{ N_{ c,\rm eff}}$ 
or $\Gamma_\alpha^{\rm spline}$ which was observed 
in \cite{Venumadhav:2015pla}.

The lepton asymmetries needed to produce the complete dark 
matter abundance ($\Omega_{ \rm s} /\Omega_{\rm DM} = 1 $) 
are quite large compared to the baryon asymmetry.
Most baryogenesis 
mechanisms produce comparable amounts
of  lepton and baryon asymmetries before 
electroweak sphaleron freeze-out.
In the  $\nu  $MSM ~\cite{Asaka:2005an}, which contains two additional  
heavier sterile neutrinos,
a larger lepton asymmetry can be produced thereafter
\cite{Shaposhnikov:2008pf,Laine:2008pg}.
However, it turns out this can  boost the lepton asymmetry by at most 
a factor 1,000  \cite{Ghiglieri:2018wbs},
and that one can reach at most
$\Omega_{ \rm s} /\Omega_{\rm DM} = 1 /10$ 
\cite{Ghiglieri:2019kbw}.
In this scenario, improving on calculations for hadronic contributions to 
active-neutrino opacities {\it  can}  be important
\cite{Ghiglieri:2019kbw}, as the lepton asymmetries are even smaller
than in fig.~\ref{f:sol1e6}.  

\subsection{Limits from Big Bang Nucleosynthesis}	\label{s:BBN}

Measuring the helium abundances as a result of BBN one obtains the upper
bound~\cite{Boyarsky:2009ix}
\begin{equation}	\label{BBN_lim}
   \frac{n_L}{s} \leq 2.5 \cdot 10^{-3}
\end{equation}
on the total lepton asymmetry.
We now compute the resulting lower limit on the mixing angle,
for which one can obtain 
$ \Omega  _ { \rm s } / \Omega  _ { \rm DM } = 1 $. 
For simplicity, we take the maximal value in \eqref{BBN_lim} as an 
initial condition at $T=4\,$GeV.
The lepton asymmetry at times prior to BBN could be higher, but 
the depletion turns out to be only on the level of a few percent for the 
low mixing angles considered here. 

We calculate, for various masses, the mixing angle that leads to the 
complete relic DM abundance. 
The results are given in table~\ref{t:thmin},
displaying the scenario where all asymmetry is in the muon flavor (left)
and the scenario where the asymmetry is split equally onto all three flavors (right).
The only non-zero neutrino Yukawa coupling is $h_\mu$. 
\begin{table}[t]  
  \caption{Mixing angle that leads to the complete relic DM abundance 
for various masses, with total initial asymmetry $n_L/s = 2.5 \cdot 10^{-3}$. 
Left: asymmetry only in the muon flavor, right: 
all three asymmetries initially equal,
    $n_{L_\alpha}/s = n_L/3s $.}
    \label{t:thmin}
  \bigskip 
  \begin{minipage}[b]{.49\linewidth}
		\centering
		\begin{tabular}{c|c}	
		$ M /\rm keV$ &  $ \sin^2(2\theta) \cdot 10^{13} $\\
		\hline 
		$ 1 $ & $ 30 $\\
		$ 2 $ & $ 11 $\\
		$ 5 $ & $ 2.8 $\\
		$ 10 $ & $ 1.03 $\\
		$ 15 $ & $ 0.61 $\\
		$ 20 $ & $ 0.39 $\\
		$ 30 $ & $ 0.23 $\\
		$ 40 $ & $ 0.17 $\\
		\end{tabular}
	
	\end{minipage}
	\begin{minipage}[b]{.49\linewidth}
		\centering
		\begin{tabular}{c|c}
		$ M /\rm keV$ &  $ \sin^2(2\theta) \cdot 10^{13} $\\
		\hline
		$ 1 $ & $ 47 $\\
		$ 2 $ & $ 16 $\\ 
		$ 5 $ & $ 4.1 $\\
		$ 10 $ & $ 1.48 $\\
		$ 15 $ & $ 0.85 $\\
		$ 20 $ & $ 0.57 $\\
		$ 30 $ & $ 0.35 $\\
		$ 40 $ & $ 0.25 $\\
		\end{tabular}
   
	\end{minipage}
\end{table}

\begin{figure}[t]
 \centering
 \includegraphics[width=.8\textwidth]{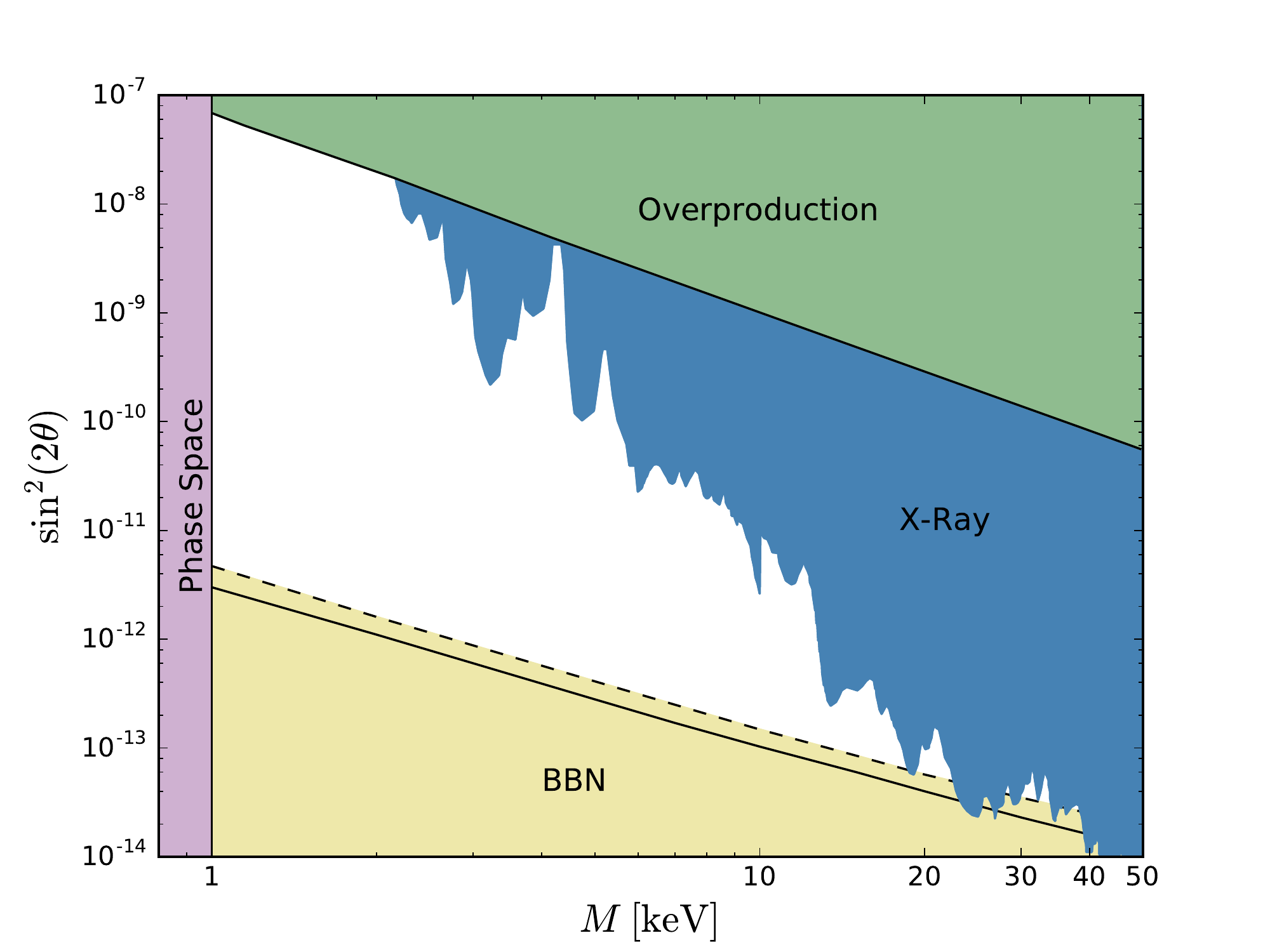}
 \caption{Combined constraints for keV sterile-neutrino dark matter. 
The X-ray constraints are taken from \cite{deGouvea:2019phk}, see 
also \cite{Essig:2013goa,Horiuchi:2013noa,Perez:2016tcq,Ng:2019gch}. 
Phase space density constraints are from \cite{Boyarsky:2008ju}. 
The BBN limit given by the solid black line holds if all of the input lepton 
asymmetry is only in the muon flavor. 
The dashed line corresponds to the BBN limit if the input lepton asymmetry 
is split equally onto all three flavors.}
 \label{f:bounds}
 \end{figure}

Combining our  lower limit  on the mixing angle  with current 
X-ray constraints
closes the parameter space for masses $M \gtrsim 40\,$keV. 
A lower mass limit of $M \gtrsim 1\,$keV can be deduced from DM phase 
space density restrictions within dense galaxies 
\cite{Tremaine:1979we,Boyarsky:2008ju}. 
In the low mass range, very high mixing angles in the Dodelson-Widrow 
scenario are excluded because they result in dark matter overproduction 
\cite{Asaka:2006nq}. 
Therefore the parameter space is bounded from all sides.
We show the combined constraints in fig.~\ref{f:bounds}.
The applicability of lower mass bounds from Milky Way satellite counts or 
Lyman-$\alpha$
methods in the case of resonant production, 
which typically leads to colder than thermal
spectra, is not clear at this stage.
Hence we did not include them in our parameter plot.
Previous calculations of such mass bounds have been performed 
e.g. in \cite{Cherry:2017dwu, Schneider:2016uqi} with the code {\it sterile-dm}.
Generally we find colder spectra than \cite{Schneider:2016uqi},
with a mean momentum that is typically between $25$\% and $50$\% lower,
depending on the region in the parameter space.

As we have seen
in sec.~\ref{s:width}, our calculations generally give larger phase 
space densities than {\it sterile-dm}, 
if it is used for resonant production ``as is'' with 1,000 default 
momentum bins.
This code was used to calculate the BBN limit in
\cite{Cherry:2017dwu,Schneider:2016uqi,Ng:2019gch,Roach:2019ctw}, 
giving much stronger limits than the ones we find, 
especially for the lower end of the mass range shown in fig.~\ref{f:bounds}.
Again, increasing the number of momentum bins, {\it sterile-dm} 
gives better agreement with our results.
On the other hand we note that our BBN limits are in closer agreement with the 
only slightly lower ones in \cite{Boyarsky:2009ix}, which are also displayed 
in 
\cite{Perez:2016tcq, Caputo:2019djj}, and also with the ones shown 
in~\cite{Boyarsky:2018tvu}, which are based on \cite{Serpico:2005bc}.

One has to keep in mind that  the BBN bound only applies to the total 
lepton asymmetry. 
In fact, $ n _ { L _ \mu  } $ could be larger than
\eqref{BBN_lim} if it is partly compensated
by the other lepton flavor asymmetries.
But the same compensation would not take place in~\eqref{cn} 
where the different flavors enter with different coefficients.
Therefore $ c _ \mu  $ would increase leading to a larger production
rate and to a weaker bound on $ \sin ^ 2 ( 2 \theta  ) $. 

\section{Summary and conclusions}	\label{s:concl}

In this work we have traced the evolution of keV sterile-neutrino phase 
space densities and lepton number densities in the early universe.
Lepton asymmetries much higher than the baryon asymmetry
significantly influence the active-neutrino spectral function and resonantly 
boost sterile-neutrino production.

Standard Model input enters our calculation in several places: 
through  susceptibilities,
which relate charges to chemical potentials, and through spectral functions 
of various currents, which determine the opacities of active neutrinos.
The sterile-neutrino dark matter production mainly happens during the 
cosmic QCD epoch when 
quarks and gluons are strongly coupled. 
Then both the hadronic susceptibilities
and hadronic spectral functions are non-perturbative. 
For hadronic susceptibilities, lattice QCD results are available and 
used in our calculation.
Lattice calculations of hadronic spectral functions 
are notoriously difficult, which in principle could lead to large theoretical
uncertainties.  
 
We have studied how different approximations for the hadronic contributions
to the active-neutrino opacity affect sterile-neutrino production. 
For initial conditions with vanishing sterile
neutrino density, and lepton asymmetries large enough so that
the produced sterile neutrinos make up all of today's dark matter,  
we find that the production is dominated by very sharp resonances 
for which the effect of the hadronic opacities are negligible.
This implies that more precise, non-perturbative determinations will 
not be needed for this scenario. 

For such large lepton asymmetries  we found much larger dark matter yields than 
previous studies. 
Therefore we could weaken lower bounds for 
the mixing angle derived from the upper bound on the total lepton asymmetry
from BBN.  
This opens up the available parameter space of this sterile-neutrino dark 
matter scenario.

\section*{Acknowledgements}
This work was supported by the Deutsche Forschungsgemeinschaft 
(DFG, German Research Foundation) -- project number 315477589 -- TRR 211.

\appendix

\section{Emergence of resonances with increasing lepton asymmetries}
\label{a} 

Here we show the negative helicity phase space densities and corresponding 
lepton asymmetries like in figs.~\ref{f:sol1e6}-\ref{f:sol3e4}, 
comparing the use of the three different opacities. 
From top to bottom we gradually increase the initial lepton asymmetry, 
whereby the resonances extend more towards higher momentum modes. 
When a mode passes through a resonance, it becomes blind to the choice 
of the active-neutrino opacity.
\begin{figure}[H]
\centering
\begin{minipage}{.4\textwidth}
	\includegraphics[width=1\textwidth]{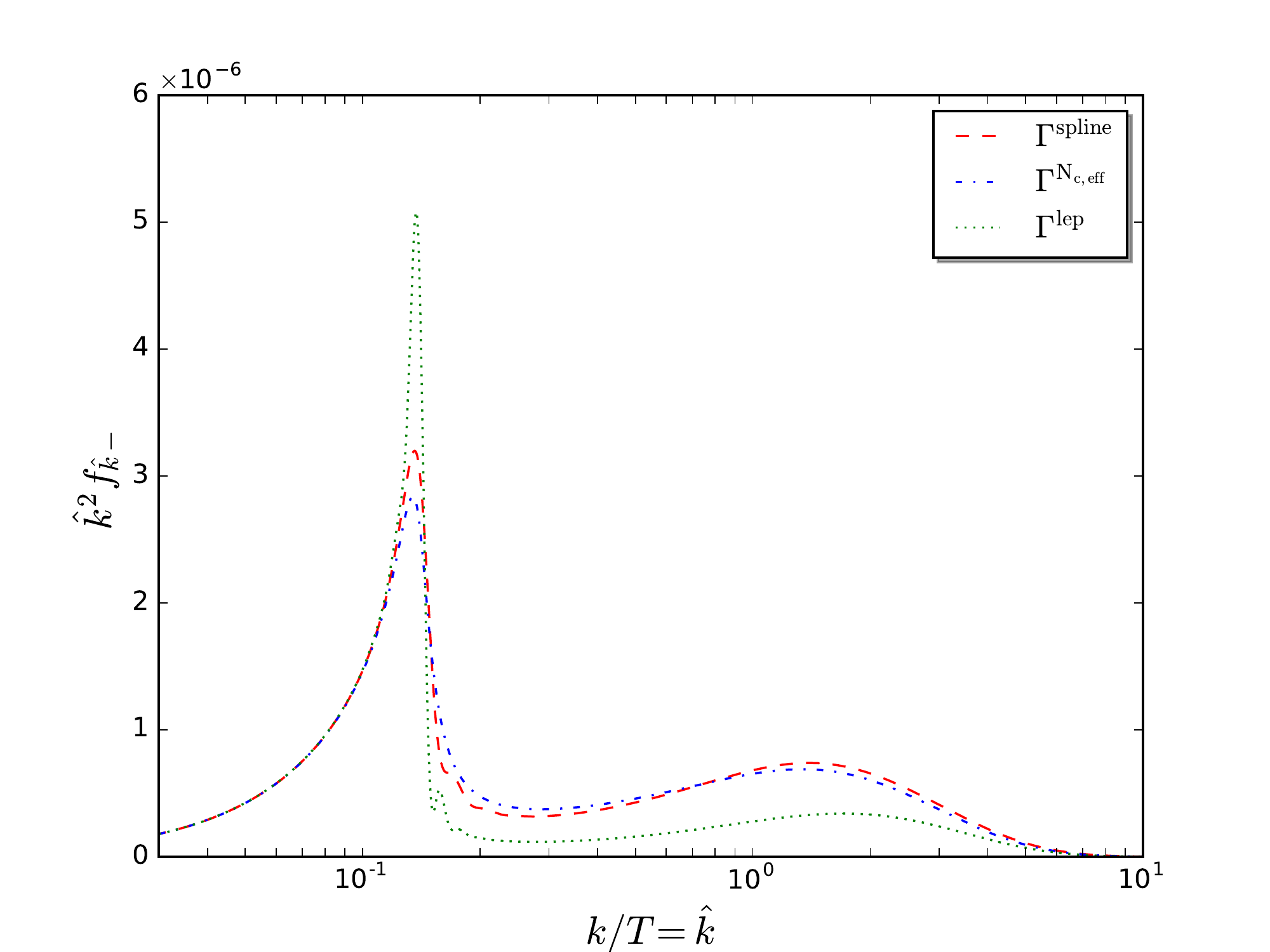}
\end{minipage}
\begin{minipage}{.4\textwidth}
	\includegraphics[width=1\textwidth]{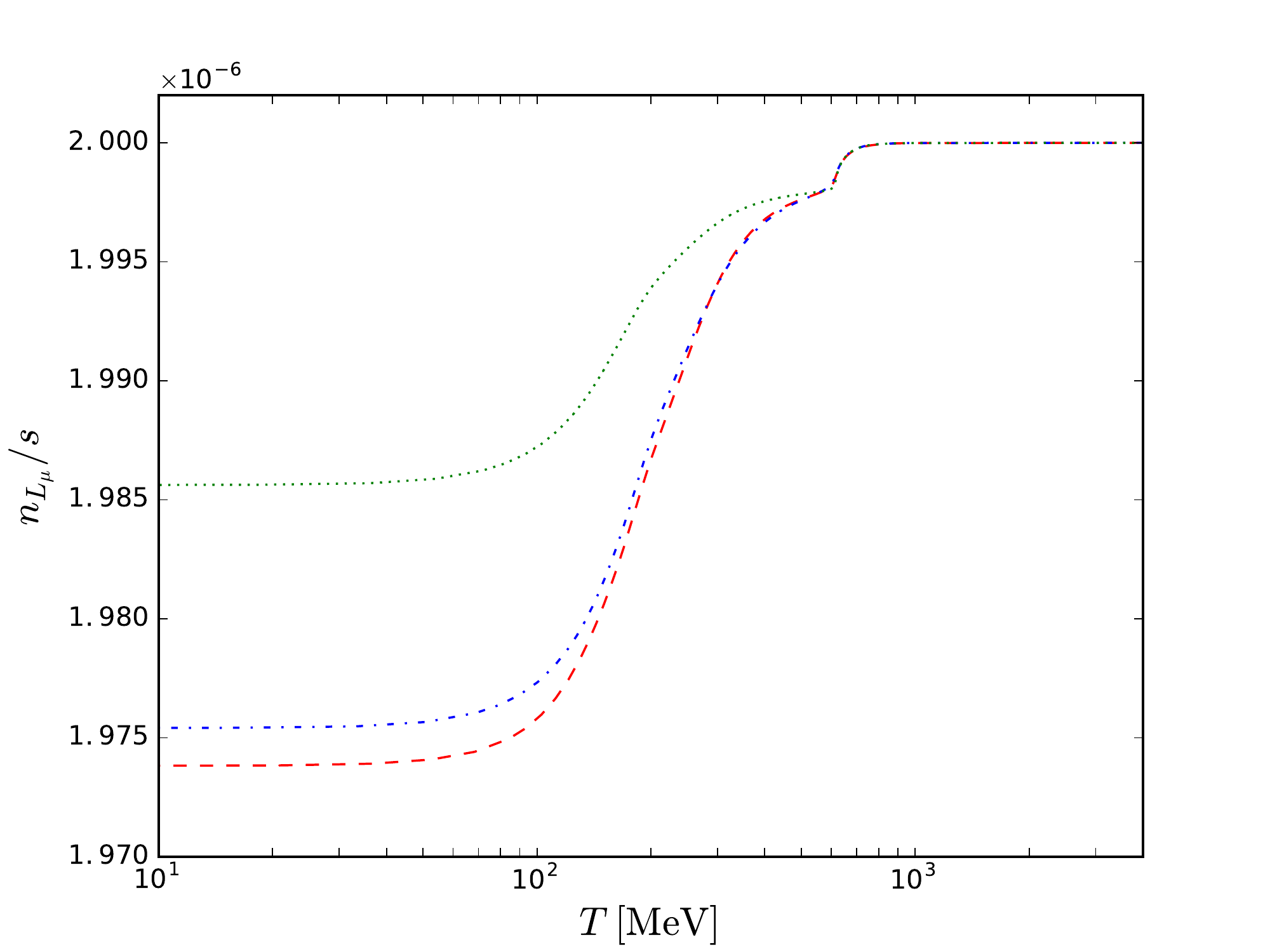}
\end{minipage}
\end{figure}

\begin{figure}[H]
\centering
\begin{minipage}{.4\textwidth}
	\includegraphics[width=\textwidth]{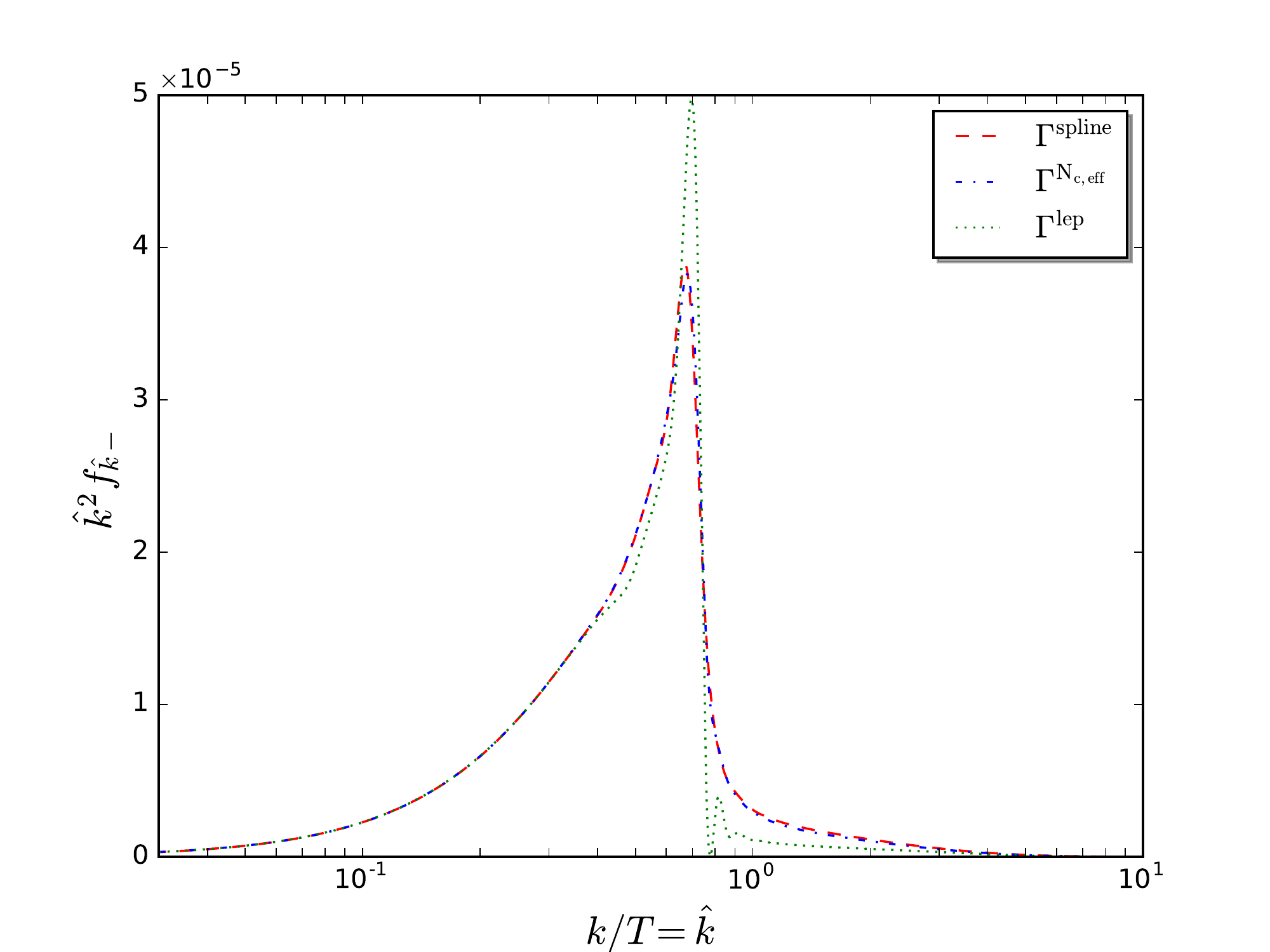}
\end{minipage}
\begin{minipage}{.4\textwidth}
	\includegraphics[width=\textwidth]{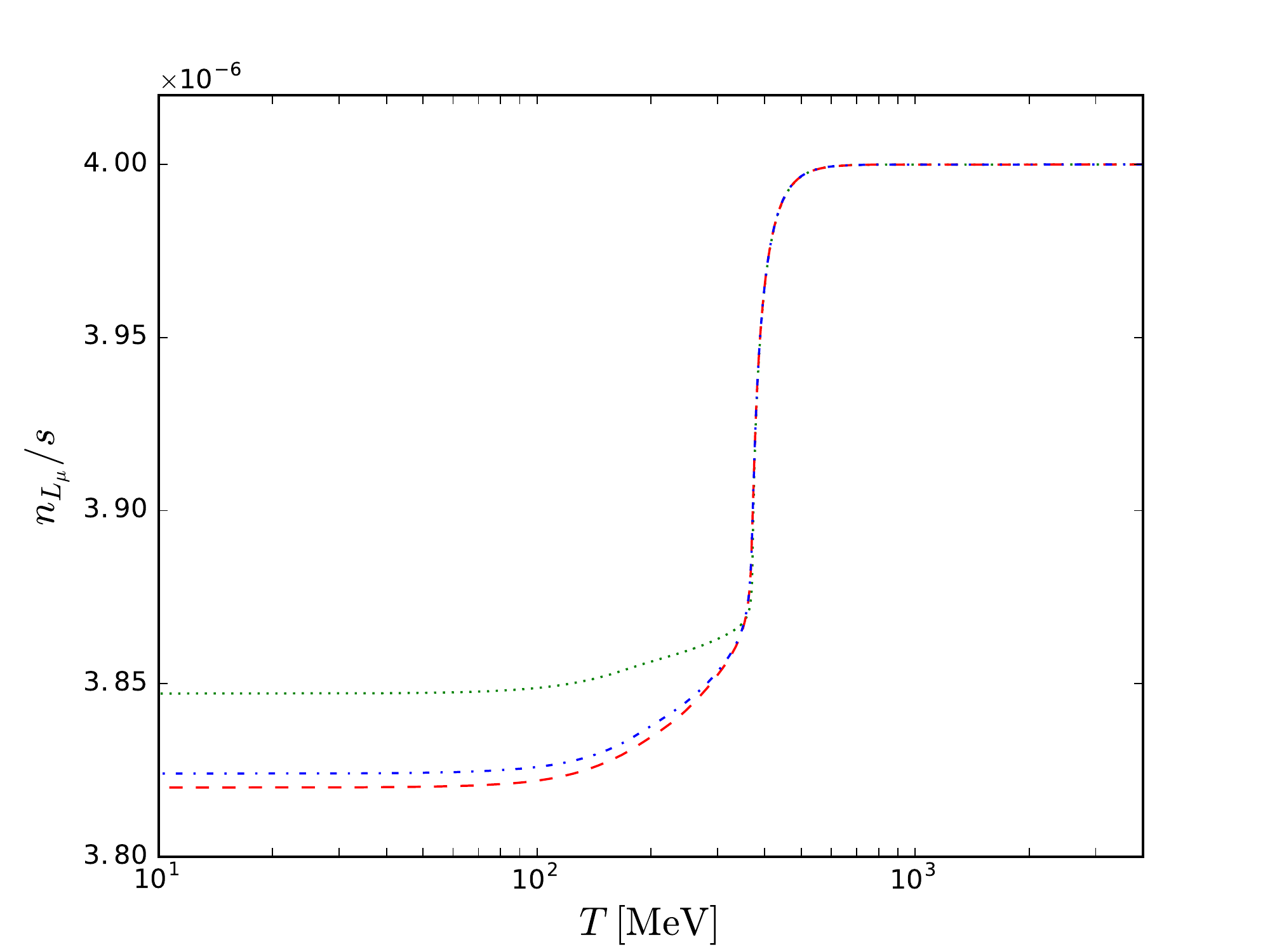}
\end{minipage}
\end{figure}

\begin{figure}[H]
\centering
\begin{minipage}{.4\textwidth}
	\includegraphics[width=\textwidth]{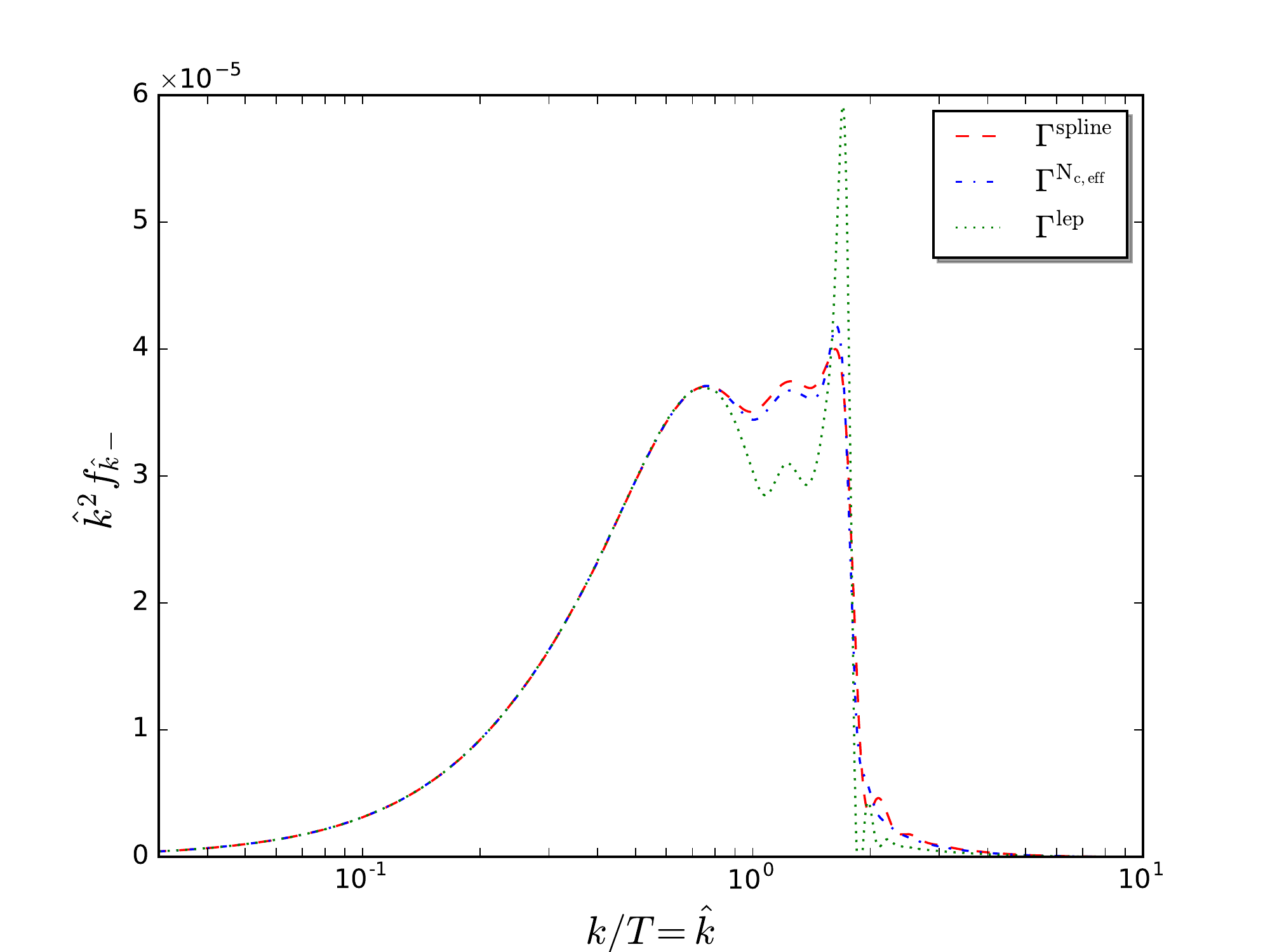}
\end{minipage}
\begin{minipage}{.4\textwidth}
	\includegraphics[width=\textwidth]{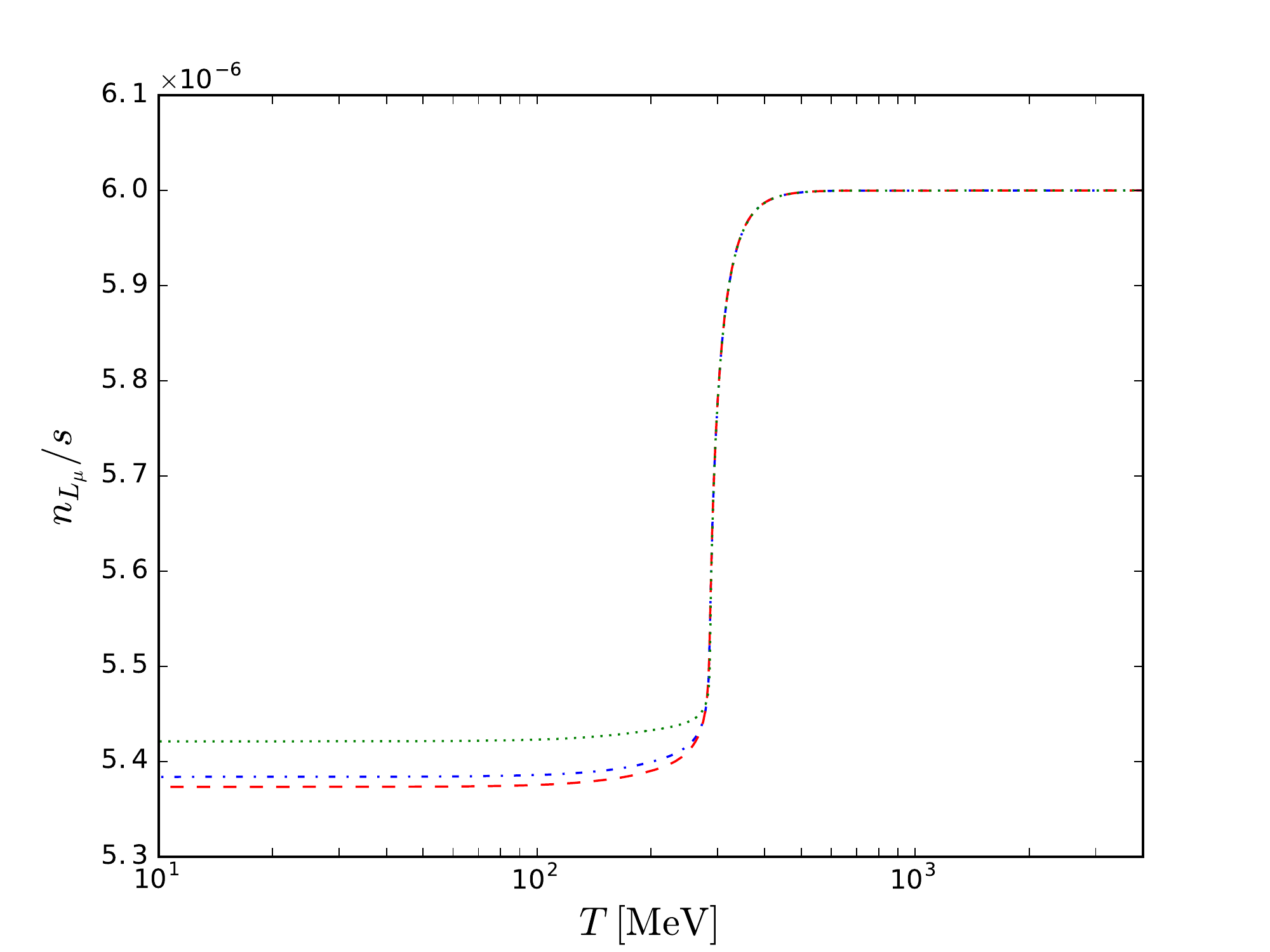}
\end{minipage}
\end{figure}

\begin{figure}[H]
\centering
\begin{minipage}{.4\textwidth}
	\includegraphics[width=\textwidth]{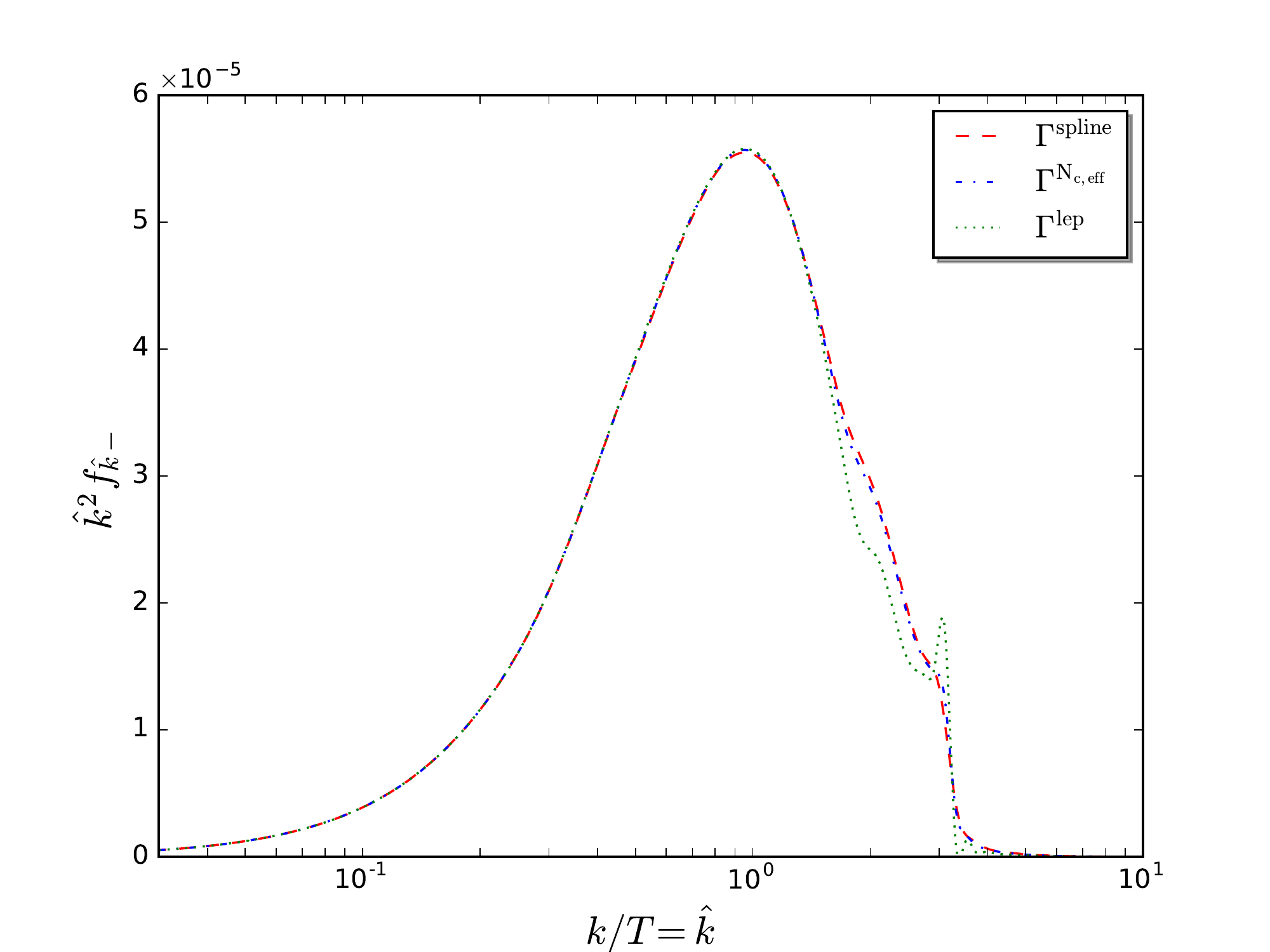}
\end{minipage}
\begin{minipage}{.4\textwidth}
	\includegraphics[width=\textwidth]{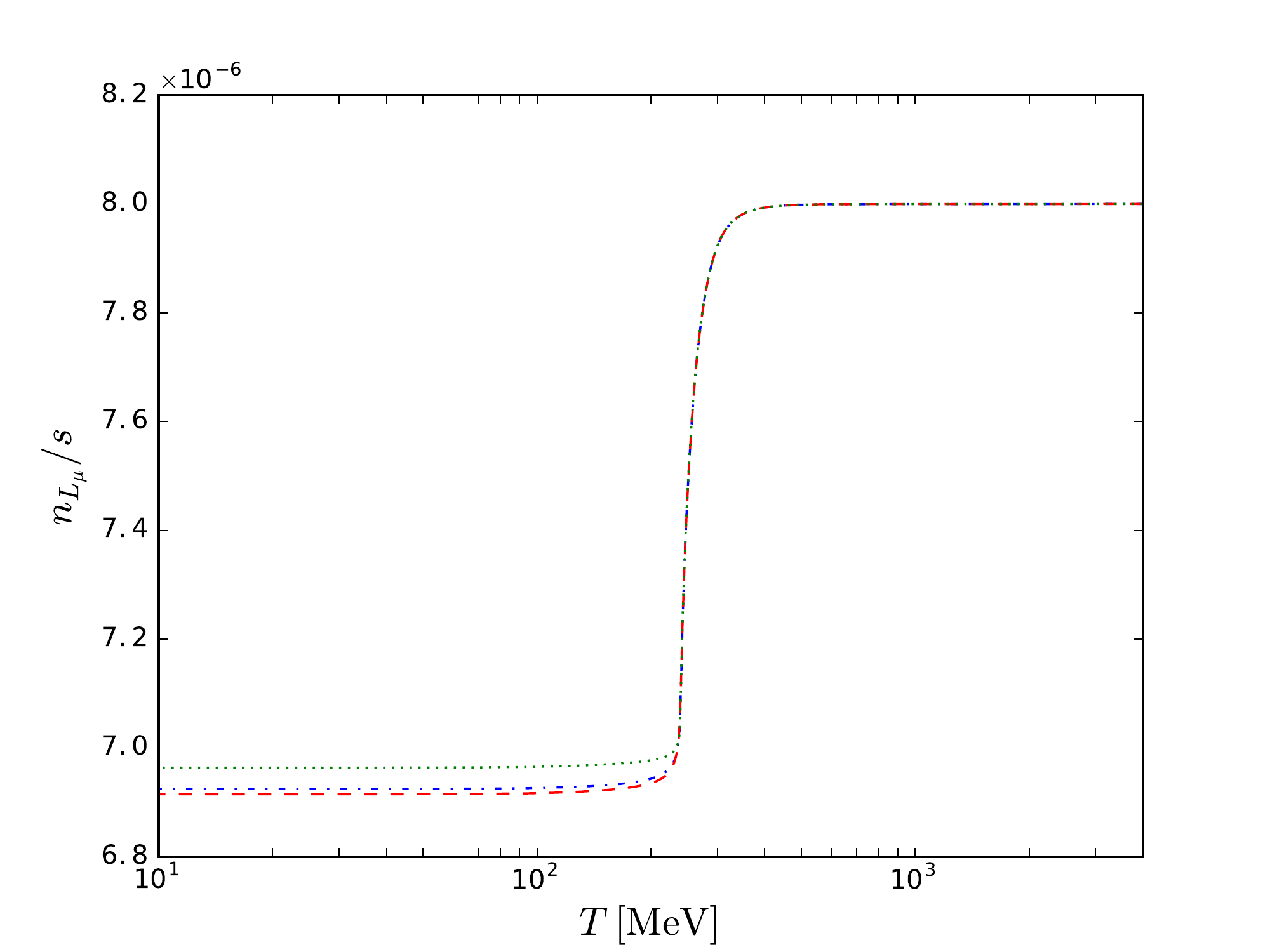}
\end{minipage}
\end{figure}

\begin{figure}[H]
\centering
\begin{minipage}{.4\textwidth}
	\includegraphics[width=\textwidth]{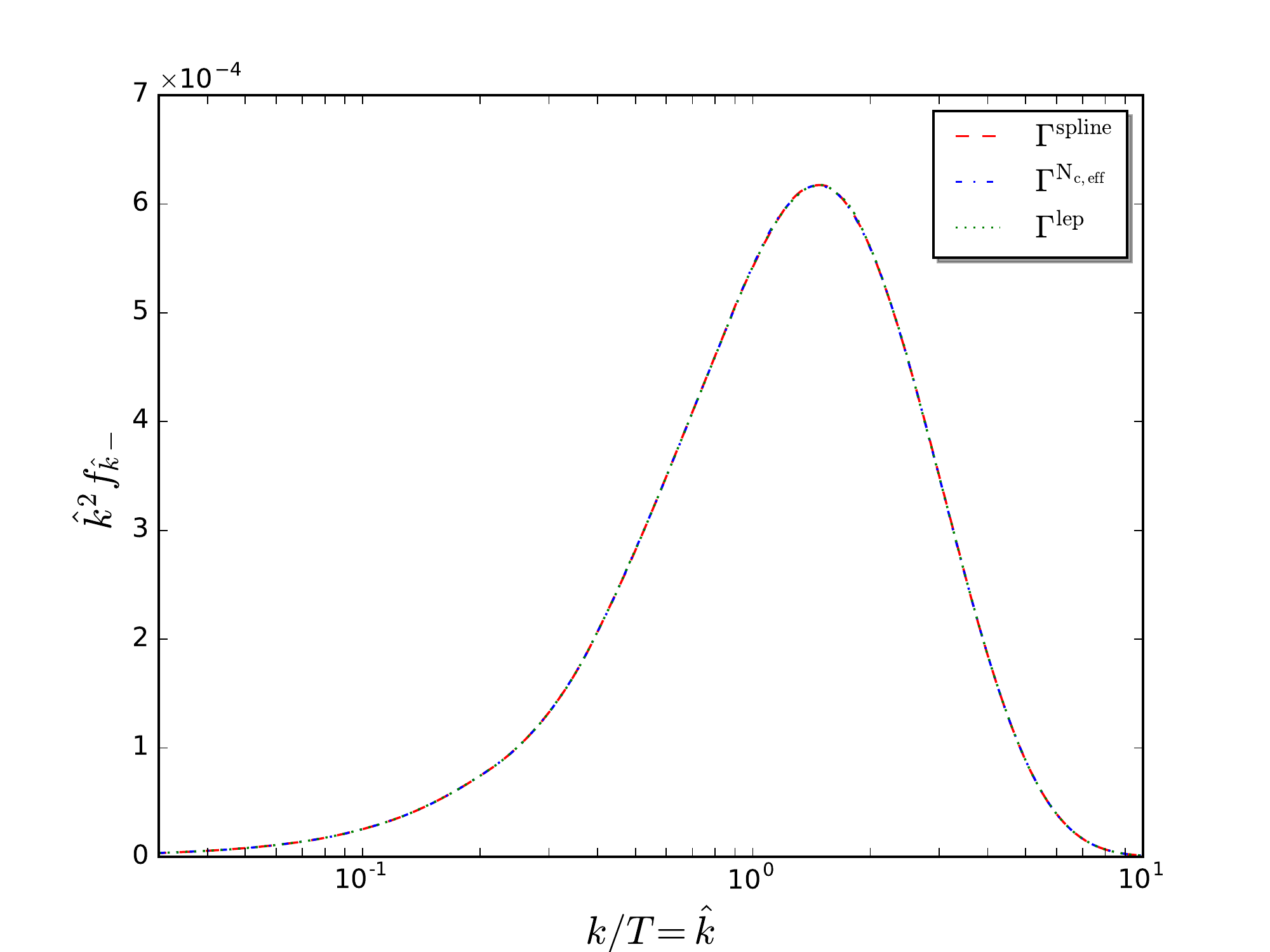}
\end{minipage}
\begin{minipage}{.4\textwidth}
	\includegraphics[width=\textwidth]{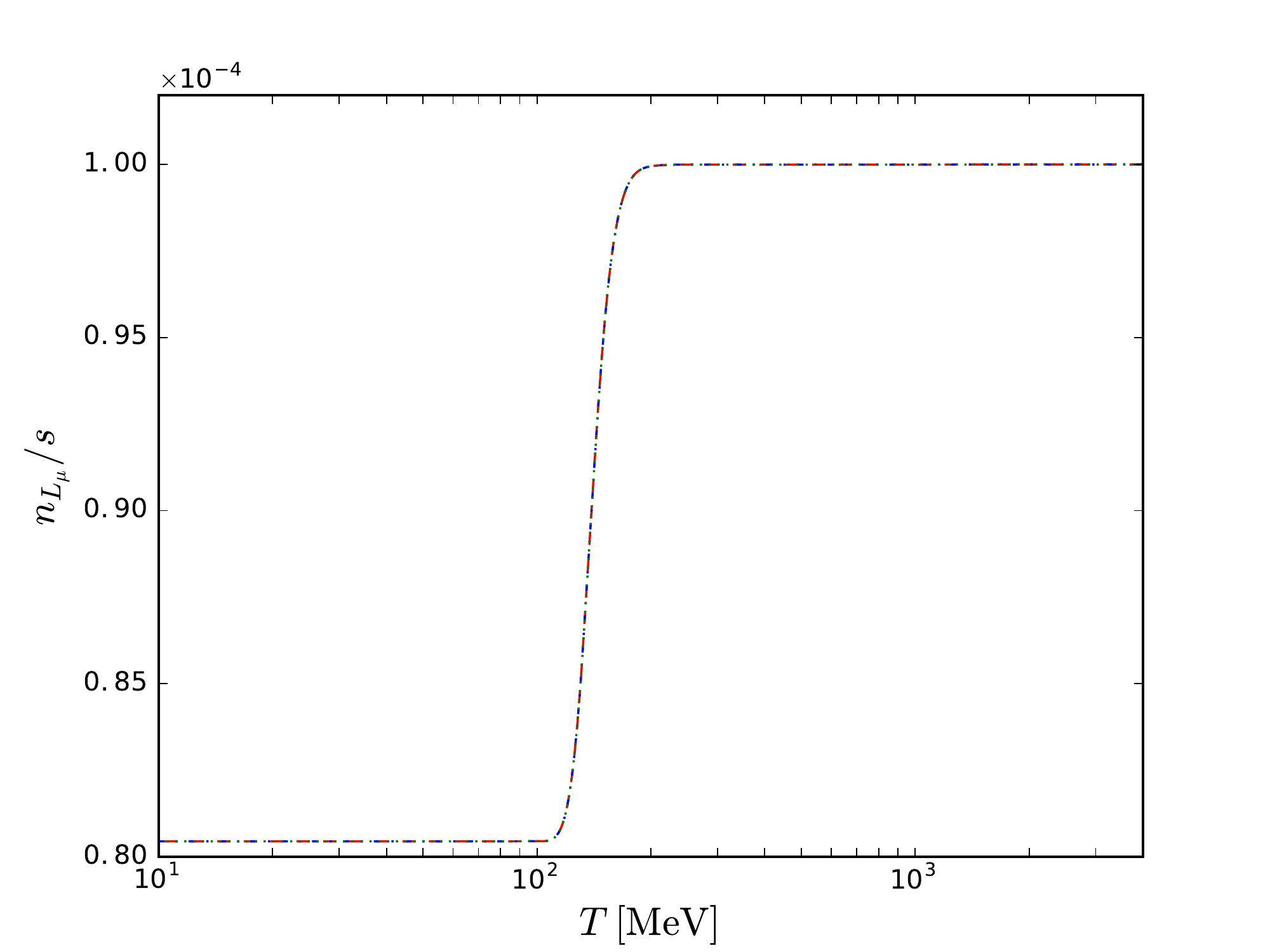}
\end{minipage}
\end{figure}

\begin{table}[H]  
  \caption{Fraction of the relic DM abundance for the initial lepton 
asymmetries used in figs.~\ref{f:sol1e6}-\ref{f:sol3e4} and in this appendix. 
We show only the abundance for the case where $\Gamma^{\rm spline}$ is used.}
    \label{t:abund}
    \bigskip
    \begin{minipage}[b]{.9\linewidth}
		\centering
		\begin{tabular}{c|c}	
		$ n_{L_\mu}/s \cdot 10^{6} $ &  $ \Omega_{ \rm s} /\Omega_{\rm DM} $\\
		\hline 
		$ 1 $ & $ 0.00057 $\\
		$ 2 $ & $ 0.00066 $\\
		$ 4 $ & $ 0.0029 $\\
		$ 6 $ & $ 0.01 $\\
		$ 8 $ & $ 0.018 $\\
		$ 10 $ & $ 0.025 $\\
		$ 100 $ & $ 0.32 $\\
		$ 350 $ & $ 1 $\\
		\end{tabular}
	\end{minipage}
\end{table}

\bibliographystyle{jhep}
\bibliography{references}
\end{document}